\documentclass[traditabstract]{aa} 
\usepackage{graphicx,dblfloatfix}
\usepackage{amsmath}
\usepackage{amssymb}
\usepackage{txfonts}
\usepackage[squaren]{SIunits} 
\usepackage{multirow}
\usepackage{dcolumn}

\newcommand{\au}{\,\hbox{au}}
\newcommand{\microns}{\micro\metre}
\usepackage{natbib}
\bibpunct{(}{)}{;}{a}{}{,}

\begin{document}

\title{Collisional modelling of the AU Microscopii debris disc}
 
\author{Ch.~Sch\"uppler\inst{1}
        \and T.~L\"ohne\inst{1} 
        \and A.~V.~Krivov\inst{1}
        \and S.~Ertel\inst{2}
        \and J.~P. Marshall\inst{3,4}        
        \and S.~Wolf\inst{5}
        \and M.~C.~Wyatt\inst{6}        
        \and J.-C. Augereau\inst{7,8}
        \and S.~A.~Metchev\inst{9}
        }
 
\institute{Astrophysikalisches Institut und Universit\"atssternwarte, 
           Friedrich-Schiller-Universit\"at Jena, 
           Schillerg\"a{\ss}chen~2--3, 07745 Jena, 
           Germany \\
\email{christian.schueppler@uni-jena.de}
 \and
European Southern Observatory, Alonso de Cordova 3107, Vitacura, Casilla 19001, Santiago, Chile 
 \and
School of Physics, University of New South Wales, NSW 2052 Sydney, Australia
 \and
Australian Centre for Astrobiology, University of New South Wales, NSW 2052 Sydney, Australia
 \and
Institut f\"ur Theoretische Physik und Astrophysik, Christian-Albrechts-Universit\"at zu Kiel, Leibnizstra\ss{}e 15, 24098 Kiel, Germany
 \and
Institute of Astronomy, University of Cambridge, Madingley Road, Cambridge CB3 0HA, UK
 \and
Univ. Grenoble Alpes, IPAG, F-38000 Grenoble, France 
 \and
CNRS, IPAG, F-38000 Grenoble, France 
 \and 
University of Western Ontario, Department of Physics and Astronomy, 1151 Richmond Avenue, London, ON N6A 3K7, Canada
 }
 
\date{Received \today}

\abstract 
{  
AU Microscopii's debris disc is among the most famous and best-studied debris 
discs, and one of only two resolved debris discs around M~stars.
We perform in depth collisional modelling of the AU Mic disc
including stellar radiative and corpuscular forces (stellar winds), 
aiming at a comprehensive understanding of the dust production and the
dust and planetesimal dynamics in the system.
Our models are compared to a suite of observational data
for thermal and scattered light emission, ranging from the ALMA radial surface
brightness profile at 1.3\,mm to spatially resolved polarisation measurements
in the visible.
Most of the data are shown to be reproduced with dust production in a belt
of planetesimals with an outer edge at around 40~au and subsequent inward transport of dust
by stellar winds.
A low dynamical excitation of the planetesimals 
with eccentricities up to $0.03$ is preferred.
The radial width of the planetesimal belt cannot be constrained tightly.
Belts that are 5\,au and 17\,au wide, 
as well as a broad 44\,au-wide belt are consistent with observations. 
All models show surface density profiles increasing
with distance from the star up to $\approx$\,40\,au, as inferred from observations.
The best model is achieved by assuming a stellar mass loss rate that exceeds 
the solar one by a factor of 50.
While the spectral energy distribution and the shape of the ALMA radial profile are well
reproduced, the models deviate from the scattered light observations more strongly.
The observations show a bluer disc colour and a lower degree of polarisation 
for projected distances $<$\,40\,au than predicted by the models. 
The problem may be mitigated by irregularly-shaped dust grains 
which have scattering properties different from the Mie spheres used in this work.
From tests with a handful of selected dust materials, we derive a preference for 
mixtures of silicate, carbon, and ice of moderate porosity. 
We also address the origin of the unresolved central excess emission
detected by ALMA and show that it cannot stem from an additional inner belt alone.
Instead, it should derive, at least partly, from the chromosphere of the
central star.

\keywords{circumstellar matter  -- 
          stars: individual: AU~Mic (GJ~803, HD~197481) -- 
          submillimetre: planetary systems -- 
          scattering --
          polarisation --
          methods: numerical}
}
\maketitle


\section{Introduction}

Debris discs are remnants of the planet formation process and consist of 
(unobservable) planetesimals and collisionally replenished dust. 
They are ubiquitous around main-sequence stars with an incidence rate of 
about 20\% for FGK~stars 
with ages from hundreds of Myr to several Gyr \citep{Eiroa2013}.
For $10-800$\,Myr-old A~stars, \cite{Su2006} quoted a higher detection rate of \mbox{$\approx$\,30\%}, 
whereas \cite{Thureau2014} found a similar frequency as for FGK-type stars.
The possible difference in the debris disc occurrence rate between the FGK and A~stars appears to be largely 
due to the observed age ranges and the typical disc decay timescale for different spectral types
\citep[e.g.,][]{Decin2003, Wyatt2007, Kains2011}.
The frequency of debris discs around M~stars remains controversial
\citep{Lestrade2006,Gautier2007,Forbrich2008,Plavchan2009,Lestrade2009,Lestrade2012}.
Despite the high abundance of M~dwarfs in the Galaxy \citep[$\sim$\,80\%,][]{Lada2006}, 
only a few discs around them have been detected so far, for which several 
reasons could exist: 
(i)~M~dwarfs have low luminosities and are older on average than
other stars so that their discs are faint and already mostly collisionally depleted.
(ii)~The blowout of dust around M~stars may be favoured since they likely possess
strong stellar winds \citep[e.g., recently][]{Johnstone2015_II,Johnstone2015_I}.
(iii)~Owing to their low stellar mass, it is easy to strip 
planetesimals from an M-dwarf disc during close encounters with massive objects.
(iv)~Debris disc surveys, including stars of various spectral types, reveal generic temperatures for warm and cold 
disc components of $\sim$\,190\,K and $\sim$\,60\,K, respectively \citep{Morales2011, Ballering2013, Chen2014}. 
If this also holds for M stars, the M-type discs would be close to the star, meaning a
faster dynamic timescale of the disc and a quick depletion of the dust reservoir.
Around a dozen of non-resolved M-star discs have been found in the recent years \citep{Forbrich2008, Plavchan2009, Chen2014}.
\cite{Theissen&West2014} found a much larger number of old M dwarfs showing mid-IR excesses, interpretable as circumstellar dust emission.
However, due to the high fractional luminosities measured, the dust is more likely attributed to planetary impacts 
within the terrestrial zone than to asteroid belt-like debris discs.
The discs around the M~stars AU~Microscopii and GJ~581 \citep{Lestrade2012} are the only ones that are spatially resolved thus far.

Since the discovery of the AU~Mic disc \citep{Kalas2004, Liu2004_science} it 
remains remarkable among the resolved debris systems in many respects.
The disc is seen edge-on with an impressive radial extent of about 150\,au and is resolved in the optical 
to near-IR \citep{Krist2005, Metchev2005, Fitzgerald2007} 
where it appears blue relative to the star.
The surface brightness profiles from the optical/near-IR images   
show shallow inner slopes at small projected distances but steepen substantially beyond 35\,au.
In addition, asymmetries on small and large scales with several local brightness maxima and minima
have been detected at stellocentric distances beyond 20\,au \citep{Liu2004_science, Krist2005, Fitzgerald2007}.
In the latest high spatial resolution STIS (Space Telescope Imaging Spectrograph) images, 
\cite{Schneider2014} found a distinct brightness enhancement above the 
disc midplane on the south-east (SE) side at about 13\,au from the star. 
Furthermore, a disc warp is discernible on the north-west (NW) side 
between 13 and 45\,au, opposed to the SE brightness bump.
There is also an NW-SE asymmetry, with the NW side brighter than the SE side interior to 20\,au.
These disc inhomogeneities hint at the existence of planetary perturbers,
causing radially localised structures such as rings, clumps, and gaps 
through planet-disc interactions \citep[e.g.,][]{Ertel&Wolf&Rodmann2012, Nesvold&Kuchner2015}.
However, no confirmation of planets in the AU~Mic system is 
found to date  \citep{Neuhaeuser2003,Masciadri2005,Metchev2005,Hebb2007}.
Alternatively, clumpy disc structures may also be due to recent 
breakups of large planetesimals \citep[e.g.,][]{Kral2015}.

The disc has been also resolved at 1.3\,mm with SMA (Submillimetre Array), 
see \cite{Wilner2012}, and with 
ALMA (Atacama Large Millimetre/submillimetre Array), see \cite{MacGregor2013}.
From the ALMA observations two distinct emission components have been identified: 
a dust belt that extends up to 40\,au and a central emission peak that remains unresolved. 
The dust belt shows an emission profile rising with the distance from the star, 
indicating a steep surface density slope. 
The central emission peak is $\approx$\,6 times brighter 
than the stellar photosphere.
\cite{MacGregor2013} surmised this emission stems from an 
inner planetesimal belt, located $\lesssim$\,3\,au from the star.
\cite{Cranmer2013} proposed it to come solely or partly
from an active stellar corona.

This great abundance of available observational data of the AU~Mic disc provides valuable 
information from the smallest up to the largest dust particles, and further
to planetesimals that produce the dust through collisional grinding.
Constraints on the properties and spatial distribution of 
(sub)micron-sized particles can be inferred from the short-wavelength observations 
where the dust's scattered light is detected. 
However in the AU~Mic edge-on system, this is impeded by the degeneracy between the scattering properties 
of grains and their supposed spatial distribution, and requires a good knowledge of the phase 
and polarisation function of small dust.
The dynamics of small particles is strongly affected by
radiative and corpuscular forces, resulting in radiative or stellar wind blowout and drag,  
depending on the grain and stellar properties.
Observations at longer wavelengths show the thermal emission of larger grains and give insights into 
their spatial distribution. 
Especially, resolved images in the radio range reveal the locations of mm-sized, 
tracing the underlying parent bodies (planetesimals).

Much work has been done in the past to characterise and model the AU~Mic system.
\cite{Augereau&Beust2006} performed a direct inversion of the visible and near-IR surface brightness profiles.
From $H$-band observations a disc surface number density was obtained that peaks around 35\,au, close 
to the location of the break in the brightness profile, which hints at a planetesimal belt around that distance. 
Furthermore, by fitting the blue disc colour they constrained the grain size distribution and found that
grains smaller than $1\,\microns$ are mandatory to explain the observations.
The study of \cite{Strubbe&Chiang2006} included the dynamics of grains for the first time.
There, a narrow birth ring of planetesimals at about 40\,au was assumed.
Micron-sized dust grains are produced through mutual collisions of the birth-ring objects.
Grains with sizes $<$\,1\,\microns{} that are smaller than the radiative and corpuscular blowout limit,  
are expelled from the system by direct radiation and stellar wind pressure. 
Slightly larger grains are either barely bound and launched into eccentric orbits, hence forming a halo, or are
transported into the inner regions of the system by Poynting-Robertson and stellar wind drag.
The largest grains follow the dynamics of their parents and remain in the birth ring.
\citeauthor{Strubbe&Chiang2006} concluded that the disc is collision dominated with an inner part devoid of 
submicron-sized grains \citep[consistent with polarisation measurements of][]{Graham2007},
and small grains mainly populate the outer part of the disc, causing  
the blue colour of the scattered light. 
\cite{Fitzgerald2007} attempted to simultaneously reproduce the scattered light profiles, the degree of polarisation, and the spectral
energy distribution (SED) with a two-zone disc, using power-law descriptions for the radial and 
size distribution of the dust in each zone.
They showed that the first zone (\mbox{35~--~40\,au}) of large particles, representative for a planetesimal ring, 
mainly accounts for the long-wavelength thermal emission, whereas the second zone (\mbox{40~--~300\,au}),
composed of small particles, well reproduces the scattered light measurements.
Recently, a dust halo was also detected in far-IR resolved \emph{Herschel} and JCMT (James Clerk Maxwell Telescope) images by Matthews et al. (in prep.). 
Their best fit supports the narrow birth-ring model of \cite{Strubbe&Chiang2006}, but is also consistent with an
extended planetesimal ring from \mbox{8~--~40\,au}, according to the model explored in \cite{Wilner2012} and \cite{MacGregor2013}.

In this study, we undertake collisional modelling
to find a generic size and radial distribution of the dust in the AU~Mic disc.
We aim at a comprehensive understanding of the dust production and the
dust and planetesimal dynamics in this system. 
For the first time, we combine constraints from scattered light and mm 
wavelength observations and search for a self-consistent collisional model that explains
all these data.
We assume an axisymmetric disc, and therefore do not 
account for the formation of substructures and asymmetries observed in the AU~Mic disc.
Section~\ref{sec:observations} gives an overview of the observational data and presents
our re-reduction of the ALMA data, resulting in a star+disc flux density and
a radial surface brightness profile at 1.3\,mm.
Section~\ref{sec:collisional_modelling} explains our collisional model.
Section~\ref{sec:outer_disc} shows the modelling of the extended, $\approx$\,40\,au-wide emission 
zone, seen in the ALMA image, which we refer to as resolved outer disc.
Section~\ref{sec:central_emission} addresses a possible origin of the unresolved central emission.
In Sect.~\ref{sec:power-law_model}, we compare our results from collisional modelling with a multidimensional power-law fitting
approach.
Conclusions are drawn in Sect.~\ref{sec:conclusions}.

 
\section{Data used}
\label{sec:observations}

\subsection{Stellar properties}
\label{sec:stellar_properties}

AU~Mic (GJ~803, HD~197481) is an M1~V dwarf at a distance of 
9.9\,pc \citep{Perryman1997, vanLeeuwen2007}.
The star is a bright X-ray and UV emitter and shows strong flaring activity \citep[e.g.,][]{Robinson2001}.
We used the photosphere model of \cite{Augereau&Beust2006}
with an effective temperature of 3700\,K, a luminosity of 0.09\,$L_\odot$, and a surface gravity
of $\log(g)=4.5$ (CGS).
We assumed a stellar mass of 0.5\,$M_\odot$, motivated by the mean of the
wide mass range given in the literature \citep[$0.3-0.6\,M_\odot$,][]{Plavchan2009,Houdebine&Doyle1994}.
Note that this assumption is roughly consistent with the photosphere model used that predicts
$\approx 0.6\,M_\odot$.

AU~Mic belongs to the $\beta$~Pic moving group (BPMG)
and its age coincides with the BPMG age by assuming 
a coeval origin of all group members.
Through the identification of the lithium depletion boundary, 
\cite{Binks&Jeffries2014} and \cite{Malo2014} found BPMG ages of 
$21\pm4$\,Myr and $26\pm3$\,Myr, respectively.
This agrees with traceback ages from \cite{Makarov2007} and
\cite{Mamajek&Bell2014}. 
The latter used  revised \emph{Hipparcos} astrometry data and
also derived an isochronal age of $22\pm3$\,Myr.

\subsection{ALMA observations}
\label{sec:ALMA_observations}

ALMA observations of AU~Mic in Band~6 (1.3\,mm) have been carried out as 
part of the cycle~0 early science observations in the context of the projects 
2011.0.00142.S (PI: D.~Wilner) and 2011.0.00274.S (PI: S.~Ertel). 
The data considered in the present work were taken on 16 June 2012 and are 
in agreement with the \cite{MacGregor2013}'s SB-4 observations.
A total of 20 operational 12-m antennae were used, spanning baselines of
\mbox{21~--~402\,m} with an effective angular resolution 
($FWHM$ of the reconstructed dirty beam) of $0.69\arcsec \times 0.79\arcsec$
and an effective field of view of $\lambda/D\approx22\arcsec$
(with $\lambda$ the observing wavelength and $D$ the single dish diameter).
\cite{MacGregor2013} presented a detailed description of the observations 
and data reduction.
Our re-reduction of the data was carried out in the Common Astronomy 
Software Applications \citep[CASA,][]{McMullin2007}. 
We followed the approach used by \cite{MacGregor2013} and came to 
consistent conclusions.

The edge-on disc has a radial extent of about 4\arcsec{}~($\approx$\,40\,au) and a position
angle of $128.41 \pm 0.13$\degr\ \citep{MacGregor2013}.
It is unresolved in its vertical direction.
We refrain from presenting the ALMA image as it would not provide information
beyond that shown in Fig.~1 of \cite{MacGregor2013}.
In order to prepare the data for our modelling, we extracted a radial 
profile along the disc major axis. 
To this end, we largely adopted the approach 
used for \emph{Herschel} data obtained in the context of the Open Time Key 
Programme DUst around NEarby Stars
\citep[DUNES,][]{Loehne2012, Eiroa2013, Ertel2014}. We first converted
the flux units in the image from Jy/beam to Jy/pixel by multiplying them with 
a factor of 0.0185 derived by integrating the core of the reconstructed dirty beam. 
Then, we rebinned the image
with an original pixel scale of 0.1\arcsec{}/pixel
to a ten times smaller pixel scale using 
a cubic spline interpolation and rotated it by an angle of 38.41\degr\ 
clockwise in order to align the disc major axis with the image $x$-axis.
From this image, we measured the radial profile along the disc 
major axis by integrating over $11\times11$ subpixel-wide boxes centred 
on the disc midplane and spaced by 74 subpixels (0.74\arcsec, about 
the $FWHM$ of the reconstructed dirty beam,
i.e. the resolution element of the image) 
left and right of the star, assumed to be located at the brightest subpixel.
We averaged the SE and NW sides of the profile to 
fit an axisymmetric model to the data.
Uncertainties were derived as a quadratic sum of 
the difference between the two sides of the profile and the background 
fluctuation measured in regions of the image where no flux is expected but 
where the sensitivity is comparable to the image centre.
As we were only interested in the dust distribution in the outer disc,
we ignored the profile point at the image centre,
which is affected by the inner, unresolved component. 
By extrapolating the disc profile from the outer points to the disc centre
and assuming that the flux measured above the extrapolated value stems from the inner
component (star and additional emission), we estimated a flux density of this component of
$0.29 \pm 0.05$\,mJy, consistent with $0.36 \pm 0.07$\,mJy in \cite{MacGregor2013}.
We estimated the uncertainty of this measurement by adding in quadrature a typical
uncertainty in the outer points (\mbox{$\sim$\,15\%}) and an absolute photometric calibration
uncertainty of 10\% \citep{MacGregor2013}.
The stellar contribution to this flux density is 0.04\,mJy,
obtained from the stellar photosphere spectrum \citep{Augereau&Beust2006} 
extrapolated to the ALMA wavelength using the Rayleigh-Jeans law.
However, this might significantly underestimate the stellar flux and the whole 
central component might originate from stellar emission \citep{Cranmer2013}.

Deviating from the complex fitting approach used by \cite{MacGregor2013} to derive the
disc parameters (including the total flux density),
we performed photometry of the disc by
integrating the flux in the rotated image in a box of $101\times21$ native image pixels.
The uncertainty was estimated from the scatter of the flux measured on eight positions above and below the disc,
where no emission is expected.
An additional calibration uncertainty 
of 10\%, consistent with \cite{MacGregor2013}, was added in quadrature. 
We found a total flux density of $8.75\pm0.98$\,mJy, which is consistent with 
$7.46\pm0.76$\,mJy from \cite{MacGregor2013}, by summing both disc components and 
including calibration uncertainty.
For the modelling, we subtracted the flux of the inner component estimated above,
so that we are left only with the flux of the disc and the star as estimated 
using our photosphere spectrum.

Note that the exact flux of the inner component has no significant impact 
on our profile or flux measurement, since we ignore the inner profile 
point affected by this component and because its contribution to the total flux density is only $0.2\sigma$.

\subsection{Auxiliary data}

In addition to the ALMA data, we considered for modelling a variety of 
photometric data from earlier work.
Furthermore, we included archival \textit{Herschel} \citep{Pilbratt2010}
PACS \citep[Photodetector Array Camera and Spectrometer,][]{Poglitsch2010} 
and SPIRE \citep[Spectral and Photometric Imaging Receiver,][]{Griffin2010, Swinyard2010}
data, obtained in the context of the debris disc Guaranteed Time Observations programme 
(PI: G.~Olofsson, see Matthews et al., in prep.).
These data were treated following the \textit{Herschel}/DUNES data 
reduction strategy \citep{Eiroa2013}. 
HIPE (\emph{Herschel} Interactive Processing Environment, \citealt{Ott2010}) 
version 11 and the calibration plan versions v56 and v11 for PACS and SPIRE, 
respectively, were used. 
PACS flux densities at $70\,\microns$ and $160\,\microns$ were measured 
with an aperture of 20\arcsec{} and annuli for the noise measurement of 
\mbox{30~--~40\arcsec{}} and \mbox{40~--~60\arcsec}, respectively. The SPIRE flux density at 
$250\,\microns$ was measured with an aperture of 30\arcsec{} and the 
noise was estimated in a region of the image where no emission from the disc 
is expected. A nearby source is visible in all \textit{Herschel} images, 
peaking at $250\,\microns$. It is well separated from the disc 
at wavelengths up to $250\,\microns$ and easily removed by a point 
source subtraction. The two sources are not well 
separated at longer wavelengths. SPIRE flux densities at $350\,\microns$ and $500\,\microns$ 
were measured using SExtractor \citep{Bertin&Arnouts1996} and a point source approximation 
(PSF photometry), since the source is unresolved at these wavelengths. 
Aperture corrections were applied to the measured fluxes. 
Instrument calibration uncertainties of 7\% for PACS 
and 6\% for SPIRE were added in quadrature to the measured uncertainty.
Note that a more detailed reduction of the \emph{Herschel} data will be given in 
Matthews et al. (in prep.).
Minor differences to the data used here exist, but all within $2\sigma$ (B.~Matthews, priv. comm.).
The differences have no impact on the results presented in this work.

We did not consider the spatially resolved information from the \emph{Herschel} images since 
the resolution is much lower than that of our ALMA image 
(e.g., $FWHM=5.8$\arcsec{} for \emph{Herschel}/PACS at 70\,\microns{} vs. $FWHM=0.74$\arcsec{} for ALMA Band~6). 
However, we inspected the PACS data for signs of bright, warm central emission 
in the form of a significantly increasing extent
of the disc from $70\,\microns$ to $160\,\microns$ \citep{Ertel2014}. 
Such a behaviour is not visible suggesting that the inner component seen 
in the ALMA images has indeed no or only a minor contribution to the fluxes 
at shorter wavelengths.
\cite{Doering2005} reported a detection of the innermost disc regions up to $\approx$\,16\,au in diameter 
via $N$-band imaging at Gemini South. 
However, only a low-level surface brightness extension along the direction of the edge-on disc was found
that has not been taken into account in our study.
Furthermore, we did not consider \emph{Spitzer}/MIPS images due to the low angular resolution
(e.g., the resolution of Spitzer at 24\,\microns{} is nearly as low as that of 
\emph{Herschel} at 70\,\microns{}). 
Any confusion, e.g. from a background galaxy, would here be difficult to disentangle from the source.

Assuming the flux density to be proportional to $\lambda^{-\delta}$ at long wavelengths, the mean 
spectral index beyond 250\,\microns{} amounts to $\delta=1.7$, close to  
the Rayleigh-Jeans (RJ) slope of a blackbody radiator ($\delta_\mathrm{RJ}=2$).
We fitted a blackbody curve to the far-IR excess 
in order to derive spectral slopes at all wavelengths. 
A temperature of 50\,K fits the data very well, consistent 
with \cite{Rebull2008} and \cite{Plavchan2009}. 
Interestingly, this is close to the generic temperature found for the outer cold components in many two-component disc systems 
\citep{Morales2011, Ballering2013, Chen2014, Pawellek2014}.
The temperature that was found for the inner disc, 190\,K, corresponds to a very small distance of 0.7\,au for a putative warm component 
in the AU~Mic system. 
This is well below the resolution limit and leaves room for further speculations about the presence of an inner unresolved disc component.
We took the nearest tabulated values for the colour corrections listed in the literature for 
the instruments used and the spectral slopes of the source (star and disc) 
derived from this fit. 
The resulting colour corrected flux densities and references 
for both the measurement and the colour correction are listed in 
Table~\ref{tab:fluxes}.

\begin{table}[htb!]
\caption{Colour-corrected photometry of AU~Mic.}
\label{tab:fluxes}
\centering
\begin{tabular}{D{.}{.}{-1}r@{\,$\pm$\,}lcp{0.5cm}c}
\hline\hline\\[-2ex]
  \multicolumn{1}{c}{Wavelength}        & \multicolumn{2}{c}{Flux density}          & Telescope/              & Ref. & Ref.   \\
  \multicolumn{1}{c}{$[\microns]$}      & \multicolumn{2}{c}{[mJy]}                 & Instrument              & Flux      & Col. corr. \\
\hline\\[-2ex]
   3.35             & 4820 & 119 & WISE                    & C12, W10  & W10         \\
   4.6              & 4260 & 119 & WISE                    & C12, W10  & W10         \\
   5.8              & 2200 & 55  & \textit{Spitzer}/IRAC   & C05       & IH          \\
   8                & 1270 & 20  & \textit{Spitzer}/IRAC   & C05       & IH          \\
   9                & 912  & 22  & Akari                   & I10       & IH          \\
  11.6              & 543  & 60  & WISE                    & C12       & W10         \\
  12                & 517  & 30  & IRAS                    & M90       & B88         \\
  18                & 246  & 36  & Akari                   & I10       & IH          \\
  22.1              & 183  & 21  & WISE                    & C12       & W10         \\
  24                & 158  & 3   & \textit{Spitzer}/MIPS   & P09       & IH          \\
  25                & 244  & 63  & IRAS                    & M90       & B88         \\
  60                & 269  & 46  & IRAS                    & M90       & B88         \\
  70                & 227  & 27  & \textit{Spitzer}/MIPS   & P09       & IH          \\
  70                & 231  & 16  & \textit{Herschel}/PACS  & HSA       & IH          \\
 100                & 680  & 149 & IRAS                    & M90       & B88         \\
 160                & 172  & 21  & \textit{Spitzer}/MIPS   & R08       & IH          \\
 160                & 243  & 17  & \textit{Herschel}/PACS  & HSA       & IH          \\
 250                & 134  & 8   & \textit{Herschel}/SPIRE & HSA       & IH          \\
 350                & 72   & 21  & CSO/SHARCII             & C05       & \dots       \\
 350                & 84.4 & 5.4 & \textit{Herschel}/SPIRE & HSA       & IH          \\
 450                & 85   & 42  & JCMT/SCUBA              & L04       & \dots       \\
 500                & 47.6 & 3.8 & \textit{Herschel}/SPIRE & HSA       & IH          \\
 850                & 14.4 & 1.8 & JCMT/SCUBA              & L04       & \dots       \\
1300                & 8.5  & 2.0 & SMA                     & W12       & \dots       \\
1300                & 8.75 & 0.98& ALMA                    & M13       & \dots       \\
\hline
\end{tabular}
\tablebib{B88: \cite{Beichman1988}, 
          C05: \cite{Chen2005}, 
          C12: \cite{Cutri2012}, 
          HSA: \textit{Herschel} Science Archive (Matthews et al., in prep.).
          IH: corresponding instrument hand book.
          I10: \cite{Ishihara2010},
          L04: \cite{Liu2004},
          M90: \cite{Moshir1990},
          M13: \cite{MacGregor2013},
          P09: \cite{Plavchan2009},
          R08: \cite{Rebull2008},
          W10: \cite{Wright2010},
          W12: \cite{Wilner2012}. 
The ALMA and \textit{Herschel} data were re-analysed in the present work.
}
\end{table}

\subsection{Scattered light}
The AU~Mic disc is 60\% brighter in $B$-band than in $I$-band relative to the star \citep{Krist2005}.
A potential explanation is a significant amount of small grains that scatter more light at shorter wavelengths. 
\cite{Augereau&Beust2006} constrained the grain size distribution by fitting the blue disc colour.
They concluded that grains down to $\sim$\,0.1\,\microns{} are necessary to explain the observations
when astronomical silicate or graphite grains are assumed.
\cite{Graham2007} obtained polarisation maps in $V$-band 
using the \emph{Hubble Space Telescope} (HST) Advanced Camera for Surveys.
They measured the degree of linear polarisation to be steeply rising from 5 to 40\% within a distance of 80\,au from the star.
The light is polarised perpendicular to the disc plane that agrees with the expected scattering behaviour by small grains with size parameters \mbox{$x \lesssim1$} 
(equivalent to typical grain sizes \mbox{$<$\,1\,\microns}). 
To model the $V$-band scattered light intensity and the degree of polarisation, 
\cite{Graham2007} assumed uniformly distributed dust between two distances from the star
and adopted the empirical \cite{Henyey&Greenstein1941} phase function  
in combination with a parameterized polarisation function. 
In their best-fit model, the particles exhibit strong forward scattering
and the innermost 40\,au of the disc are devoid of grains.
Later on, \cite{Shen2009} showed that the results from \cite{Graham2007} 
can be reproduced by a distribution of sphere cluster aggregates with size distribution index $-$3.5, volume-equivalent radii
between 0.1 and 0.4\,\microns, and a porosity of 60\%.

In this work, our models are compared with the optical and near-IR surface brightness profiles 
from \cite{Fitzgerald2007} and \cite{Schneider2014}
and with the measured degree of polarisation from \cite{Graham2007}.
As our modelling deals with axisymmetric discs, all observed profiles had been  
averaged over the SE and NW disc sides in order to remove the observed surface brightness anomalies.

\subsection{Gas in the disc}

Some studies searched for circumstellar gas in the AU~Mic disc. 
From a non-detection of submillimetre CO emission \cite{Liu2004} inferred an H$_2$ mass of $\leq$\,1.3\,$M_\oplus$. 
\cite{Roberge2005} reduced the H$_2$ mass limit to $0.07\,M_\oplus$ due to a non-detection of \mbox{far-UV} H$_2$ absorption.
\cite{France2007} analysed fluorescent H$_2$ emission and found a total gas mass between 
\mbox{$4\times10^{-4}\,M_\oplus$} and \mbox{$6\times10^{-6}\,M_\oplus$}, but stated
that this detection might also come from a cloud that extends beyond the disc.
From X-ray spectroscopy \cite{Schneider&Schmitt2010} reported a maximum
H column density of \mbox{$\sim10^{19}\,\centi\metre^{-2}$}, which is about five times 
higher than the interstellar value.
However, since observational data point towards a very low gas content in the AU~Mic disc, 
we neglect the influence of any gas in our investigations.

\section{Collisional modelling}
\label{sec:collisional_modelling}

Collisional modelling with our code \texttt{ACE} (Analysis of Collisional Evolution) and its application to debris disc systems 
is described extensively in previous 
papers \citep[][among others]{Krivov2006,Krivov2008,Mueller2010,Reidemeister2011,Loehne2012,Krivov2013,Schueppler2014}. 
The code assumes a disc of planetesimals orbiting a central star,
simulates their collisional grinding, and follows the collisional and dynamical evolution of the system under a variety of physical effects.
\texttt{ACE} works with a three-dimensional logarithmic grid for objects mass, pericentric distance, and 
eccentricity. 
The simulations consider a number of parameters, including those that define the initial distribution of planetesimals, 
the strength of particle transport, and material properties such as the critical fragmentation energy. 
The following sections detail the parameters chosen in our modelling.

\subsection{Stellar winds and mass loss rate}
\label{sec:sw_strength+mass_loss}

Stellar winds (SW) are expected to play a crucial role in discs around late-type stars such as AU Mic 
where they are typically stronger than for early-type stars
\citep[e.g.,][]{Plavchan2005, Plavchan2009, Strubbe&Chiang2006, Reidemeister2011, Schueppler2014}.
The total stellar wind force that acts on the circumstellar dust 
can be decomposed to direct stellar wind pressure, which is directed radially outwards, and stellar wind drag, which works
like a headwind on the particles that causes them to gradually lose orbital energy and angular
momentum, and therefore, to spiral inwards \citep{Burns1979}.
Depending on the stellar wind strengths and the size of the particles they are interacting with, 
the result could be either a grain transport towards the star or a blowout of smaller grains from the system.

Previous studies proposed various stellar winds strengths for the AU~Mic system, expressed  
in terms of multiples of the solar mass loss rate, $\dot{M}_\odot=2\times10^{-14}\,M_\odot~{\rm yr}^{-1}$. 
\cite{Strubbe&Chiang2006} constrained the stellar mass loss rate to be \mbox{$\dot{M}_\star<10\,\dot{M}_\odot$} which 
was in good agreement with the observed SED and the \mbox{$V-H$} colour profile of the disc. 
This relatively weak wind strength means that the disc is rather 
dominated by destructive grain-grain collisions than transport processes.
\cite{Plavchan2009} estimated \mbox{$\dot{M}_\star<50\,\dot{M}_\odot$} from considerations of \emph{Spitzer}/IRS observations.
\cite{Augereau&Beust2006} took into account the active nature of AU~Mic, and
found \mbox{$\dot{M}_\star \approx 50\,\dot{M}_\odot$} during quiescent states
and \mbox{$\dot{M}_\star\approx 2500\,\dot{M}_\odot$} during flares. 
The star spends about 10\% of the time in flare phases with typical flare durations of several minutes \citep{Pagano2000}.
Averaging over quiescent and flare phases yields a mean stellar mass loss of $\dot{M}_\star = 300\,\dot{M}_\odot$.
This high wind strength means that the direct stellar wind pressure is larger than the radiation pressure which 
causes an efficient transport of small dust into outer disc regions.

In our modelling, we considered moderate and strong wind strengths by assuming
50\,$\dot{M}_\odot$ and 300\,$\dot{M}_\odot$. 
Throughout this paper, we refer to both cases as SW50 and SW300. 
According to \cite{Augereau&Beust2006}, SW50 is a typical model where the flares are ignored, 
while SW300 is an attempt to account for the impact of the episodic stellar activity.
In all simulations involving stellar winds, 
we used Eq.~(7) of \cite{Gustafson1994} for the total stellar wind force acting on a particle and assumed 
a wind speed \mbox{$v_\mathrm{sw}=400\,\kilo\metre~ \second^{-1}$}
\citep{Strubbe&Chiang2006}.

\subsection{Grain material}
\label{sec:grain_material}

\cite{Graham2007} and \cite{Fitzgerald2007} found evidence for 
porous aggregates containing silicate, carbon, and ice.
Guided by these results, we used porous grains composed of  
astronomical silicate \citep{Draine2003b}, ACH2 amorphous carbon \citep{Zubko1996}, 
and water ice \citep{Li1998}.
Since the ACH2 refractive data from \cite{Zubko1996} terminate at
\mbox{$\lambda\approx1$\,mm} and do not cover the full wavelength range 
for all available photometric measurements, 
we extrapolated the real and imaginary parts up to \mbox{$\lambda=2$\,mm}
by the function $c_1~(\lambda/1\,\milli\metre)^{c_2}$. 
The extrapolation gives $(c_1,c_2)=(16.2,0.27)$ and $(c_1,c_2)=(5.6,-0.26)$ 
for real and imaginary parts, respectively.

We generated different porous silicate-carbon-ice mixtures and calculated 
the refractive indexes of the composites by applying the \cite{Bruggeman1935} mixing rule.
The materials considered in this study are given in Table~\ref{tab:materials}.
For brevity, we use the identifiers M1 to M5 when referring to the different compositions. 

\renewcommand{\arraystretch}{1.3}
\begin{table}[h!]
\centering
\caption{Materials used for modelling.}
\label{tab:materials}
\begin{tabular}{p{0.5cm}cccc}
 \hline\hline
 Name   &  Composition                &  Density                               & \multicolumn{2}{c}{Blowout size $[\microns]$}  \\
        &                             &  $[\gram\,\centi\metre^{-3}]$          &  SW50          &  SW300           \\
 \hline
 M1     &  sil33+car33+vac33          &  1.78                                  &  0.07          &  0.48            \\
 M2     &  sil50+car50                &  2.68                                  &  0.04          &  0.35 \\
 M3     &  sil25+car25+ice25+vac25    &  1.64                                  &  0.07          &  0.51 \\
 M4     &  sil10+car10+ice40+vac40    &  1.02                                  &  0.11          &  0.71 \\
 M5     &  sil15+car15+vac70          &  0.80                                  &  0.16          &  0.93 \\
 \hline
\end{tabular}
\tablefoot{
 Numbers after material names denote their volume fraction in percent, e.g.
 sil25+car25+ice25+vac25 = 25\%~silicates + 25\%~carbon + 25\%~ice + 25\%~vacuum.
 The bulk densities of the composites are given by $\rho_\mathrm{mix}=\sum \rho_i ~\sigma_i$,
 where $\sigma_i$ denotes the volume fraction of the species, and 
 $\rho_i$ their densities 
 ($\rho_\mathrm{sil}=3.5\,\gram\,\centi\metre^{-3}$, $\rho_\mathrm{car}=1.85\,\gram\,\centi\metre^{-3}$, 
 $\rho_\mathrm{ice}=1.2\,\gram\,\centi\metre^{-3}$).
 Blowout sizes are grain radii for which $\beta$ equals 0.5 (see text).
}
\end{table}

\begin{figure}[t!]
 \resizebox{\hsize}{!}{\includegraphics{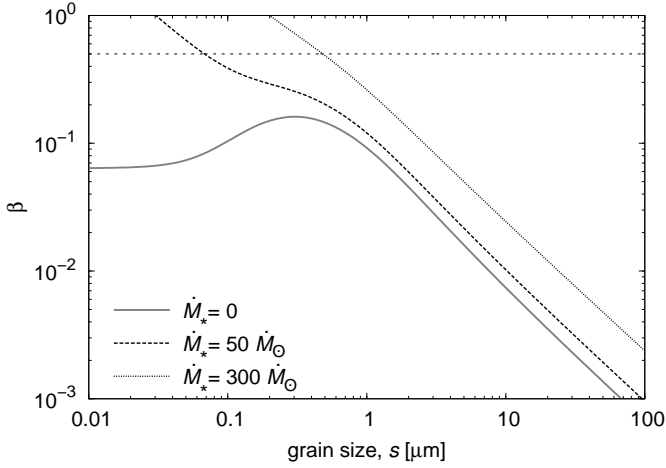}}
 \caption{$\beta$-ratio for different assumptions of AU~Mic's stellar mass loss rate $\dot{M}_\star$.  
 The material consists of 33\% carbon, 33\% silicate, and 
 33\% vacuum (M1). 
 The horizontal dashed line depicts $\beta=0.5$.
 Particles with $\beta>0.5$ are ejected 
 from the system owing to radiation and stellar wind pressure.}
 \label{fig:beta}
\end{figure}

We assumed spherical dust grains. 
Each grain has a certain radius $s$ which is denoted as its size.
Optical properties of the grains were determined by means of Mie theory \citep{Bohren1983}.
Figure~\ref{fig:beta} shows \mbox{$\beta=\beta_\mathrm{rad}+\beta_\mathrm{sw}=F_\mathrm{rad}/F_\mathrm{grav}+F_\mathrm{sw}/F_\mathrm{grav}$}, 
the ratio of direct radiation and direct stellar wind pressure forces to stellar gravity \citep{Burns1979} as a function of grain size.
A detailed description of $\beta_\mathrm{sw}$ includes a dependency on the distance from the star, $r$,
due to radial variations of the stellar wind velocity. 
However, this effect is weak for \mbox{$r>15$\,au} \citep[][]{Augereau&Beust2006} and we considered $\beta_\mathrm{sw}$ independent of $r$.
Further, we assumed the reflection of wind particles on the surface of the dust grains 
that means an efficiency of momentum coupling of \mbox{$C_\text{D}=2$} \citep{Gustafson1994}.

The orbits of the collisionally-produced dust grains become more and more eccentric with increasing $\beta$.
For \mbox{$\beta>0.5$}, the grains get eccentricities $>$\,1 and are expelled from the system on hyperbolic orbits
if they are released from bodies on circular orbits.
As a measure for the blowout size, we define $s_\mathrm{blow}=s(\beta=0.5)$.
However, when particles are released from eccentric orbits, 
the blowout occurs  between $\beta=(1-e)/2$ at the periastron and $(1+e)/2$ at the apastron 
\citep{Burns1979}, where $e$ is the typical eccentricity of 
the parent bodies that eject the dust.
\texttt{ACE} computes the orbital elements of the fragments 
depending on the masses, semi-major axes, eccentricities, 
and $\beta$'s of the colliding bodies, and thus, automatically accounts for bound and unbound grains
launched from non-circular orbits \citep{Krivov2006}.
Therefore, the minimum size of \emph{bound} grains in the debris disc system is  
slightly smaller than $s_\mathrm{blow}$. 
Nevertheless, we found this effect to be small and $s_\mathrm{blow}$ 
can be considered a representative value for the minimum grain size.

In the case of weak or no stellar winds, $\beta$ is smaller than the blowout limit 0.5 
for all grain sizes (see Fig.~\ref{fig:beta}, $\dot{M}_\star=0$). 
With increasing stellar wind strength the dust particles experience a stronger radial wind pressure.
Assuming a radiation pressure efficiency $Q_\mathrm{pr}=1$ \citep{Burns1979}, yields
$F_\mathrm{sw}/F_\mathrm{rad}\sim0.5$ for SW50 and $F_\mathrm{sw}/F_\mathrm{rad}\sim3$ for SW300.
As a result, if the stellar winds are sufficiently strong,
the $\beta$-ratio that would otherwise be smaller than 0.5 can
exceed this critical value at the smallest dust sizes.
This defines a blowout size, terminating the lower end of the size distribution (Table~\ref{tab:materials}). 
Note that Fig.~\ref{fig:beta} shows $\beta$ for M1 exemplarily.  
The other materials given in Table~\ref{tab:materials} follow a similar dependence of $\beta$ on size.

The mechanical strength of the materials was defined 
by a modified version of the \cite{Benz1999} description 
of the critical energy for fragmentation and dispersal, 
\begin{align}
 Q_\text{D}^\star=\left[Q_\text{D,s}\left(\frac{s}{1\,\text{m}}\right)^{b_\text{s}}+
 Q_\text{D,g}\left(\frac{s}{1\,\text{km}}\right)^{b_\text{g}}\right]
 \left(\frac{v_\text{imp}}{3\,\text{km}~\text{s}^{-1}}\right)^{\kappa},
\end{align}
where ($Q_\text{D,s}$, $b_\text{s}$) and ($Q_\text{D,g}$, $b_\text{g}$) 
are two pairs of constants, characterising the strength and gravity regime, 
respectively, and $v_\text{imp}$ denotes the impact speed. 
$Q_\text{D}^\star$ is the impact energy per unit target mass 
necessary to disperse the target into fragments where the largest 
fragment contains at most half of the original target mass.
For objects smaller than 100\,m, $Q_\text{D}^\star$ is mainly determined by 
the material strength, 
while for larger sizes the gravitational binding energy dominates. 
The material strength also depends on the material composition, 
but this dependence is still insufficiently constrained by laboratory work.
Therefore, we used the same $Q_\text{D}^\star$ for all materials 
given in Table~\ref{tab:materials} and assumed    
\mbox{$Q_\text{D,s}=Q_\text{D,g}=5\times10^{6}$\,erg~g$^{-1}$}, \mbox{$b_\text{s}=-0.37$}, 
\mbox{$b_\text{g}=1.38$}, and \mbox{$\kappa=0.5$} \citep[][and references therein]{Loehne2012}.

\begin{table*}[htb!]
\newcolumntype{C}[1]{>{\centering}m{#1}}
\centering
\caption{Description of \texttt{ACE} runs.}
\label{tab:run_description}
\begin{tabular}{C{2.2cm}|cccccC{1.1cm}cccC{1.5cm}c}
 \hline \hline
 Disc  type                        & Extent$^\mathrm{a)}$         & $r$-distr.$^\mathrm{b)}$      &  $\dot{M}_\star^{\phantom{\star}\mathrm{c)}}$   & $e_\mathrm{max}^\mathrm{\phantom{max}d)}$ & $i_\mathrm{max}^\mathrm{\phantom{max}e)}$ & Material$^\mathrm{f)}$ & $M_\mathrm{d}^\mathrm{\phantom{d}g)}$ & $s_\mathrm{max}^{\prime\mathrm{\phantom{max}h)}}$ & $M_\mathrm{min}^\mathrm{\phantom{min}i)}$ &  Grid resol.$^\mathrm{j)}$ & Sect.$^\mathrm{k)}$        \\
                                   & [au]                         &                               &  [$\dot{M}_\odot$]       &                                           &            [rad]                          &                        & $[M_\oplus]$                          & [m]                                               & $[M_\oplus]$                              &                            &                            \\
 \hline 
 \multirow{9}{*}{narrow outer~PB}  &\multirow{3}{*}{37.5 -- 42.5} & \multirow{3}{*}{$\propto r^0$}&  0                  & \multirow{3}{*}{0.03}                     & \multirow{3}{*}{0.015}                    & \multirow{3}{*}{M1}    & $1.5\times10^{-3}$                    & 20                                                & 0.02                                      & \multirow{3}{*}{m1,e1,p1}  &\multirow{3}{*}{\ref{sec:sw}}\\
                                   &                              &                               &  50                 &                                           &                                           &                        & $1.7\times10^{-3}$                    & 100                                               & 0.05                                      &                            &\\
                                   &                              &                               &  300                &                                           &                                           &                        & $2.1\times10^{-3}$                    & 250                                               & 0.1                                       &                            &\\
 \cline{2-12}
                                   & \multirow{3}{*}{37.5 -- 42.5}& \multirow{3}{*}{$\propto r^0$}&  \multirow{3}{*}{50}& \multirow{3}{*}{0.03}                     & \multirow{3}{*}{0.015}                    & M2                     & $2.4\times10^{-3}$                    & 100                                               & 0.06                                      & \multirow{3}{*}{m1,e1,p1}  &\multirow{3}{*}{\ref{sec:material_composition}}\\
                                   &                              &                               &                     &                                           &                                           & M3                     & $1.5\times10^{-3}$                    & 100                                               & 0.05                                      &                            &                                           \\
                                   &                              &                               &                     &                                           &                                           & M4                     & $1.4\times10^{-3}$                    & 200                                               & 0.05                                      &                            &                                           \\
 \cline{2-12}
                                   & \multirow{2}{*}{37.5 -- 42.5}& \multirow{2}{*}{$\propto r^0$}&  \multirow{2}{*}{50}& 0.01                                      & 0.005                                     & \multirow{2}{*}{M1}    & $1.7\times10^{-3}$                    &  30                                               & 0.05                                      &\multirow{2}{*}{m1,e1,p1}   &\multirow{2}{*}{\ref{sec:excitation}}      \\
                                   &                              &                               &                     & 0.1                                       & 0.05                                      &                        & $6.4\times10^{-4}$                    & 100                                               & 0.03                                      &                            &                                           \\
 \hline 
 \multirow{2}{*}{extended outer~PB}& 25.5 -- 42.5                 & $\propto r^0$                 &  \multirow{2}{*}{50}& \multirow{2}{*}{0.03}                     & \multirow{2}{*}{0.015}                    & \multirow{2}{*}{M1}    & $1.7\times10^{-3}$                    & 100                                               & 0.08                                      & m1,e1,p1                   &\multirow{2}{*}{\ref{sec:belt_width}}      \\
                                   & 1.0 -- 45                    & $\propto r^{-1.5}$            &                     &                                           &                                           &                        & $2.1\times10^{-3}$                    & 100                                               & 0.2                                      & m2,e1,p1                   &                                           \\
 \hline
 \multirow{3}{*}{narrow inner~PB}  & \multirow{3}{*}{2.8 -- 3.2}  & \multirow{3}{*}{$\propto r^0$}&  \multirow{3}{*}{50}& \multirow{3}{*}{0.03}                     & \multirow{3}{*}{0.015}                    & M1                     & $6.2\times10^{-6}$                    & $1\times10^5$                                     & 0.02                                      & \multirow{3}{*}{m1,e1,p2}  &\multirow{3}{*}{\ref{sec:central_emission}}\\
                                   &                              &                               &                     &                                           &                                           & M3                     & $5.4\times10^{-6}$                    & $4\times10^5$                                     & 0.02                                      &\\
                                   &                              &                               &                     &                                           &                                           & M5                     & $4.5\times10^{-6}$                    & $4\times10^5$                                     & 0.02                                      &\\
 \hline     
\end{tabular}
\tablefoot{%
$^\mathrm{a)}$Radial extent of the planetesimal belt; 
$^\mathrm{b)}$Specification of the radial distribution of planetesimals; 
$^\mathrm{c)}$Stellar mass loss rate in units of the solar value;
$^\mathrm{d)}$Maximum eccentricity of the planetesimal orbits;
$^\mathrm{e)}$Semi-opening angle of the planetesimal disc;
$^\mathrm{f)}$Name of dust material used (symbol explanation in Table~\ref{tab:materials});
$^\mathrm{g)}$Derived dust mass (grains up to 1\,mm in radius); 
$^\mathrm{h)}$Object size for which the collisional lifetime equals the physical age of the AU~Mic system;  
$^\mathrm{i)}$Estimated minimum total disc mass (grains up to size $s^\prime_\mathrm{max}$);
$^\mathrm{j)}$Specification of grid: 
37 bins in mass from $5\times10^{-18}$ to $1\times10^{10}\,\gram$ (m1), 
40 bins in mass from $8\times10^{-18}$ to $1\times10^{13}\,\gram$ (m2), 
22 bins in eccentricity from 0.001 to 2 (e2), 
30 bins in pericentric distance from 1 to 60\,au (p1),
20 bins in pericentric distance from 0.1 to 7\,au (p2);
$^\mathrm{k)}$Section number in this work where the runs are addressed.}
\end{table*}

\subsection{Setup}
Table~\ref{tab:run_description} lists all \texttt{ACE} simulations discussed in this paper.
We distinguish between three disc types, named by the width and location of their planetesimal belt (PB): 
(i)~a narrow outer PB; (ii)~a radially extended PB, and
(iii)~a narrow inner PB. 
Configurations
(i) and (ii) describe models for the outer resolved disc,
whereas 
(iii) corresponds to a model for the central emission seen by ALMA.
In all simulations, we assumed the discs to be composed of planetesimals up to at least 10\,m in radius with an 
initial size distribution $\propto s^{-3.5}$.
The planetesimals are large enough to ensure that their collisional lifetime is longer than the time needed 
to reach a quasi-steady state material distribution at dust sizes \citep{Loehne2008}.
Thus, the results of the simulations are independent of the maximum size of the objects.
The code also accounts for cratering and bouncing collisions 
where the bulk of one or both colliders stays 
intact but fragments are released. 
For each collision, the distribution of fragments is assumed to follow a power law with a mass
distribution index $-$1.83 down to the smallest objects represented in the grid.
The planetesimal discs were assumed to be stirred to a maximum eccentricity, $e_\mathrm{max}$.
The bodies in the disc were assumed to be uniformly distributed within a 
semi-opening angle \mbox{$i_\mathrm{max}=e_\mathrm{max}/2$} (energy equipartition).
After reaching a quasi-steady state, each simulation 
was evolved until the dust mass had reduced sufficiently to reproduce the observed excess emission.
In general, the simulation time, $t_\mathrm{sim}$, 
does not coincide with the physical age, \mbox{$t_\mathrm{phys}\approx20-30$\,Myr}, of the AU~Mic system.
For all models except the one with planetesimals between 1 and 45\,au, 
$t_\mathrm{sim}$ was 2 to 4 times shorter than $t_\mathrm{phys}$.
These times are more than enough to bring the model discs to collisional equilibrium.
The initial conditions of our model discs are, however, fiducial.
Different combinations of total initial mass and initial size distribution can lead to the same observed dust configuration, 
though after different times.
An estimate for the minimal initial disc mass necessary to provide the observed amount of dust after $t_\mathrm{phys}$ 
can be obtained by assuming that the collisional lifetimes, $t_\mathrm{col}$, of the largest bodies equal the system age,
enabling them to sustain sufficient dust production over that time. 
The required sizes for the largest bodies, $s^\prime_\mathrm{max}$, typically exceed those in our simulation runs.
We extrapolated the size distributions and collisional lifetimes to derive $s^\prime_\mathrm{max}$ such that 
$t_\mathrm{col}(s_\mathrm{max}^\prime) \approx t_\mathrm{phys}$, and then, 
determined the minimum initial disc masses $M_\mathrm{min}$ for particle sizes up to $s^\prime_\mathrm{max}$.
The power-law distributions in the m-to-km size range facilitate that extrapolation.
The results are listed in Table~\ref{tab:run_description}. 
These masses reflect the minimum disc masses necessary to achieve simulation times 
that would approach $t_\mathrm{phys}$.
Larger planetesimals and potential exo-Plutos cannot be assessed 
this way because their direct contributions to the collisional cascade are negligible.


\section{Resolved outer disc}
\label{sec:outer_disc}

First, we seek to explain the available observational data on the resolved outer AU~Mic disc, including the SED, 
the ALMA radial brightness profile, and scattered light measurements.  
In the following, we present several modelling attempts, illustrating
the influence of different parameters and their compatibility with the data.
We start with an investigation of the stellar wind strength to find an appropriate reference run (Sect.~\ref{sec:sw}).
In a subsequent analysis (Sects.~\ref{sec:material_composition}, \ref{sec:excitation}, and \ref{sec:belt_width}), 
we vary other initial disc parameters of this reference run in order to more tightly 
constrain the parameters and reveal degeneracies among them.

\begin{figure*}[htb!]
\centering
\includegraphics[width=17cm]{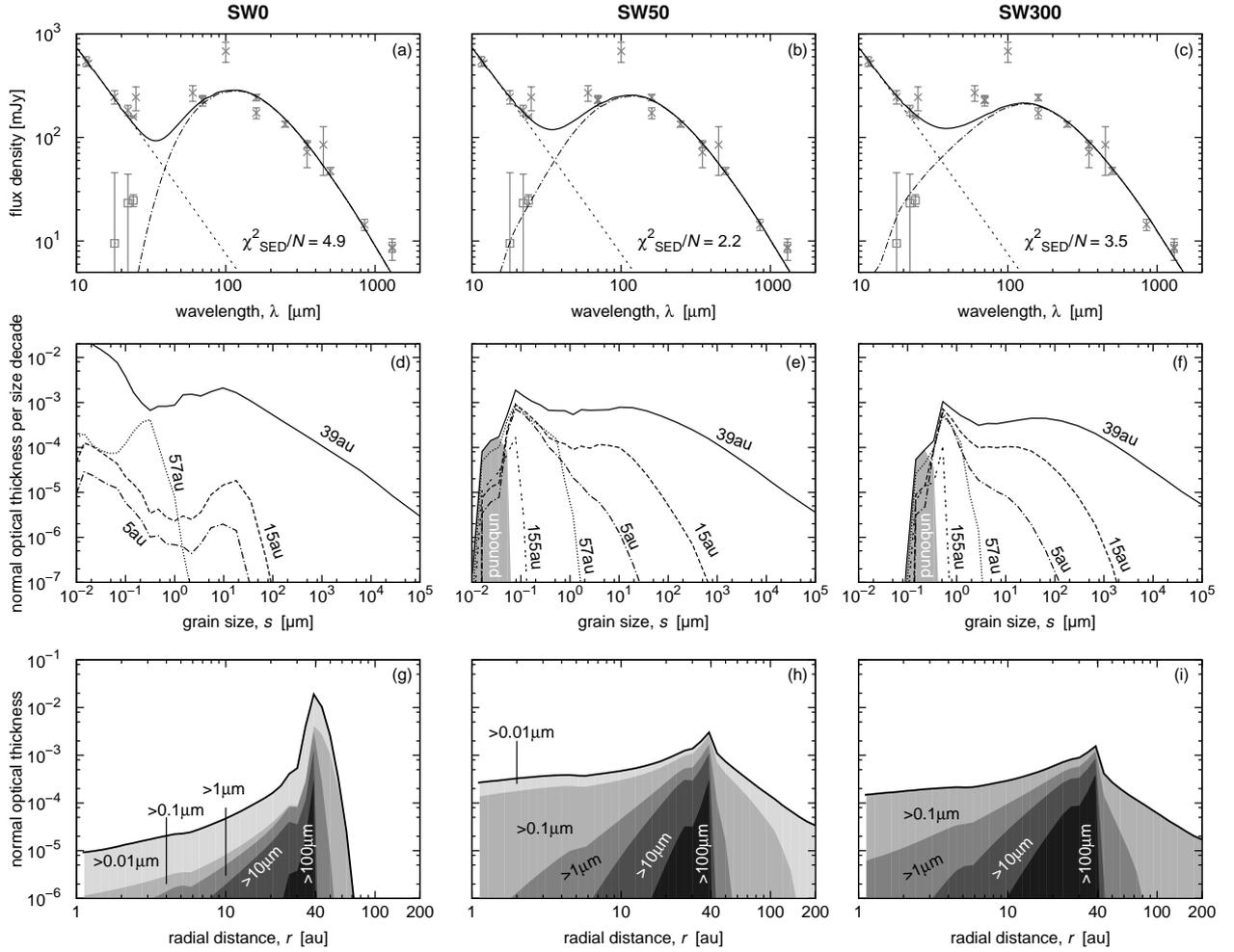}
 \caption{Disc models of narrow planetesimal belts centred at 40\,au with 
 different assumptions for the stellar wind (SW) strength (\emph{from left to right}
 \mbox{$\dot{M}_\star = 0, 50, 300\,\dot{M}_\odot$}).
 \emph{(a) -- (c):} SEDs. Crosses show photometric data and squares star-subtracted photometric data.
 The dashed lines depict the stellar photosphere, the dash-dotted lines the disc emission, and the solid lines the star+disc emission. 
 The $\chi_\mathrm{SED}^2/N$ were obtained using the \mbox{$N=20$} photometric points in the plotted range.
 \emph{(d) -- (f):} normal optical thickness as a function of grain size at different distances from the star.  
 The solid curves (\mbox{$r=39$\,au}) show the particle distribution 
 within the parent belt. Shaded regions depict the contribution of grains on unbound orbits at \mbox{$r=39$\,au}.
 \emph{(g) -- (i):} normal optical thickness as a function of distance from the star. Shaded regions
 mark the contributions from different size ranges. 
 }
 \label{fig:NR_SEDs_tau}
\end{figure*}

\subsection{Stellar wind strength}
\label{sec:sw}

In a first step, we sought after a reasonable assumption for the stellar wind strength
and carried out the runs SW0, SW50, and SW300.
SW0 does not assume any stellar wind activity at all (\mbox{$\dot{M}_\star=0$}).
Here, the transport of particles is controlled by Poynting-Robertson drag only.
In SW50 and SW300, moderate and extreme stellar winds are considered, 
given by 50 times and 300 times the solar strength, respectively 
(see Sect.~\ref{sec:sw_strength+mass_loss} for justification of these values).

For other disc parameters, we took reasonable values that have been preferred by previous studies.
Inspired by the birth-ring scenario of \cite{Strubbe&Chiang2006}, we assumed a radially narrow planetesimal belt 
centred at a stellocentric distance \mbox{$r_\mathrm{PB}=40$\,au}, having the full width \mbox{$\Delta r_\mathrm{PB}=5$\,au}. 
In all three runs, the maximum eccentricity of the planetesimals was set to \mbox{$e_\text{max}=0.03$}, 
motivated by \cite{Loehne2012} and \cite{Schueppler2014} who showed a preference for low dynamical excitations 
in debris discs around late-type stars.
For the grain material, we chose a homogeneous mixture with volume filled in equal parts by
astronomical silicate, amorphous carbon, and vacuum, denoted as M1 (Table~\ref{tab:materials}).
Other assumptions for the material composition, 
$e_\mathrm{max}$, and $\Delta r_\mathrm{PB}$ are probed 
in the Sects.~\ref{sec:material_composition}, \ref{sec:excitation}, and \ref{sec:belt_width}.

\subsubsection{Impact on thermal emission}
\label{sec:thermal_emission}

Figure~\ref{fig:NR_SEDs_tau} shows SEDs and normal optical thicknesses, $\tau_\perp$, for SW0, SW50, and SW300.
The vertical height of each SED was fitted by searching for the timestep in the \texttt{ACE} simulations 
where the dust mass has an appropriate value to reproduce the observed level of thermal emission.
For the SEDs, we determined $\chi^2_\mathrm{SED} /N $, which is the sum of the squares of the deviations of our models 
from $N$ individual photometric points, divided by $N$. 
To this end, we considered the \mbox{$N=20$} SED points for $\lambda>10\,\microns$ (Table~\ref{tab:fluxes}).
We stress that the models are not fitted to the data through variations of the initial disc parameters.
The $\chi^2_\mathrm{SED} /N $ metric merely serves for a better comparison of the SED models.
All models show good fidelity with the photometry between $160-850\,\microns{}$ but markedly 
underestimate or tend to underestimate the 1.3\,mm data.
At shorter wavelengths, SW50 provides a good match to the data, whereas SW0 significantly underestimates the mid-IR points
and SW300 the 70\,\microns{} point.

In the planetesimal zone of run SW0, the amount of particles continuously increases with decreasing grain size apart
from an underabundance around \mbox{$s=0.4$\,\microns{}} (Fig.~\ref{fig:NR_SEDs_tau}d, \mbox{$r=39$\,au}).
Although the smallest grains in the range of a few tens of microns are the most frequent, they barely affect the SED 
as their thermal emission is negligible in the IR.
Thus, the effective minimum grain size is given by the maximum of the size distribution at around 10\,\microns{}. 
This large grain size is characteristic for discs with low dynamical excitation of the planetesimals 
\citep{Thebault2008, Loehne2012, Krivov2013, Schueppler2014}.
The size distributions in SW50 and SW300 show maxima near 0.07\,\microns{} and 0.5\,\microns{}, respectively (Fig.~\ref{fig:NR_SEDs_tau}e,f).
These values are close to the blowout sizes, $s_\mathrm{blow}$ (Table~\ref{tab:materials}).
Most of the smaller grains move on unbound orbits and leave the system on short timescales.
Towards larger grains, the size distributions in the parent belt 
are nearly flat until they become significantly steeper at sizes between 10 and 100\,\microns.
This behaviour is caused by the strong transport due to stellar wind drag, which favours
smaller particles and quickly removes them from the planetesimal ring 
\citep[e.g.,][]{Vitense2010,Reidemeister2011,Wyatt2011,Krivov2013}.
As a result, the normal optical thickness is radially uniform for \mbox{$r<r_\mathrm{PB}$} in SW50 and SW300
(Fig.~\ref{fig:NR_SEDs_tau}h,i).

Direct stellar radiation pressure, aided by stellar wind pressure in SW50 and SW300,
pushes submicron-sized, barely bound grains to eccentric orbits resulting in extended dust halos.
However, both because of low dynamical excitation \citep{Thebault2008} and fast transport \citep{Strubbe&Chiang2006}, 
the halos are tenuous and exhibit radial profiles with outer slopes that are significantly steeper
than the $-$1.5 predicted for ``canonical'' debris discs \citep{Krivov2006,Strubbe&Chiang2006}.

For all three runs, the surface mass density of solids, $\Sigma$, is rising with 
distance up to $r_\mathrm{PB}$ and shows a bow-like shape in a log-log \mbox{$\Sigma-r$} diagram.
Averaged over \mbox{$r=1-30$\,au}, we measured \mbox{$\Sigma \propto r^{2.0}$} for SW0, and 
\mbox{$\Sigma \propto r^{2.7}$} for SW50 and SW300.
The latter is close to $r^{2.8}$, derived by \cite{MacGregor2013}. 
The dust mass (objects with radii up to \mbox{$s=1$\,mm}) is about $2\times10^{-3}\,M_\oplus$, 
consistent with previous estimates \citep{Augereau&Beust2006}. 
The three models have a fractional luminosity of \mbox{$4-5\times10^{-4}$}, 
which is not far from $6\times10^{-4}$ obtained 
by a modified blackbody fit \citep{Liu2004}.

\begin{figure}[t!]
 \resizebox{\hsize}{!}{\includegraphics{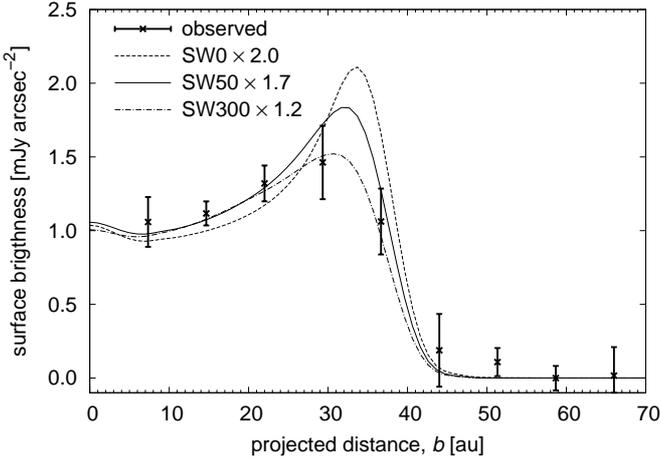}}
 \caption{ALMA radial surface brightness profile at 1.3\,mm (dots) vs. models (lines).
 Synthetic profiles are multiplied by the factors given in the legend 
 to compensate imperfections of the modelling procedure (explained in the text). 
 }
 \label{fig:NR_ALMA_profile}
\end{figure}%

Figure~\ref{fig:NR_ALMA_profile} shows the surface brightness profile extracted from the ALMA 1.3\,mm image.
The profile increases with distance up to $\approx\,$30\,au and drops steeply beyond.  
Since long-wavelength observations reveal the spatial distribution of large particles  
that have similar orbital elements as the planetesimals, 
this break indicates the outer edge of a dust-producing planetesimal zone.
For modelling of this profile, we assumed a disc inclination of 89.5$^\circ$ from face-on \citep{Krist2005} and 
convolved the synthetic 1.3\,mm images for SW0, SW50, and SW300 with the ALMA
reconstructed dirty beam as produced by CASA after proper rotation. 
From these images, we extracted the radial profiles by integrating over 
a 0.1\arcsec{} central strip along the disc major axis.  
We allow for vertical scaling of the gathered synthetic profiles.
The scaling factors account for an imperfect reproduction of the total flux density at 1.3\,mm,
visible by comparing the models with the observed SED.
Such imperfections stem, e.g., from uncertainties in the absolute flux calibration or
in the optical properties of the dust grains \citep{Loehne2012}.
Further, they derive from a real inability of our models 
to reproduce the ALMA flux density along with the other photometric data. 
We surmise that the latter reason mainly causes the models to distinctly underestimate the  
ALMA flux density and necessitates high scaling factors (Fig.~\ref{fig:NR_SEDs_tau}). 
In addition, the dust model used possibly underestimates the dust emissivity at mm wavelengths.
Figure~\ref{fig:NR_ALMA_profile} clearly shows that 
the surface brightness peak moves to smaller projected distances with increasing stellar wind strength, reflecting 
the enhanced inward transport of particles.

\begin{figure*}[t!]
 \centering
 \includegraphics[width=18cm]{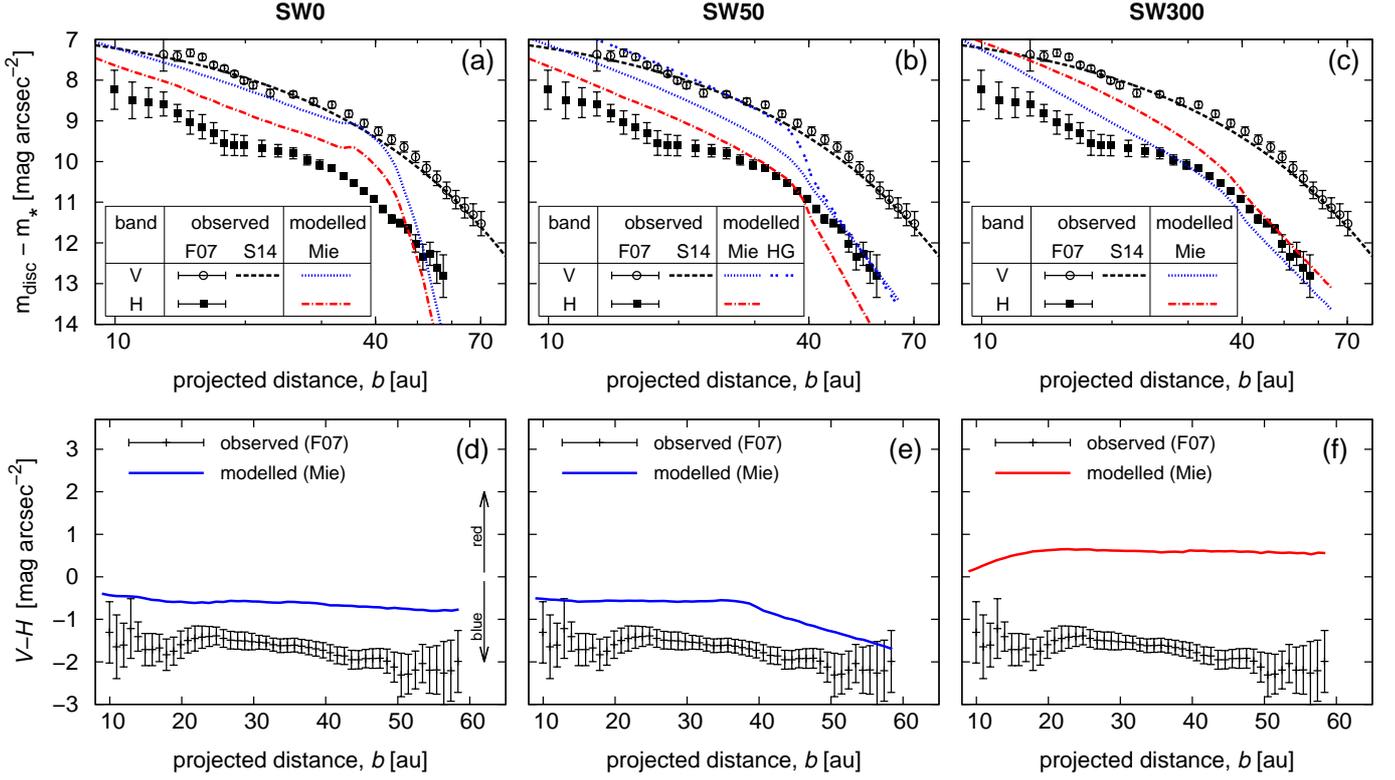}
 \caption{Scattered light profiles of the AU~Mic disc.
  \emph{(a) -- (c):} blue and red lines show synthetic midplane surface brightness profiles 
 in $V$- and $H$-band relative to the stellar brightness
 for the disc models SW0, SW50, and SW300 (Sect.~\ref{sec:sw}).
 Curves labelled with Mie in the legend were calculated with scattering phase functions obtained by means of Mie theory.
 The blue double-dotted line in panel \emph{(b)} depicts the $V$-band profile generated with the best-fit Henyey-Greenstein (HG) 
 phase function model of \cite{Graham2007}.
 Dots represent measurements with $1\sigma$ errors from \cite{Fitzgerald2007}, abbreviated as F07,
 averaged over the SE and NW wings (circles for $V$-band, squares for $H$-band). 
 The black dashed line shows the averaged $V$-band fit to both wings obtained by \cite{Schneider2014}, abbreviated as S14. 
 \emph{(d) -- (f):} observed and modelled colour profiles, $V-H$, obtained with the F07 data and the Mie profiles 
 shown in panels \emph{(a) -- (c)}. 
 }
 \label{fig:scattered_light}
\end{figure*}

\subsubsection{Impact on scattered light}
\label{sec:scattered_light}

We calculated the surface brightness of scattered light for our models along the projected distance, $b$, in the disc midplane by
\begin{align}
 S(b)=L_{\star}\int\int\frac{\pi s^2 Q_{\text{sca}}\, S_{11}(\phi)\, n(r,s)}{4\pi r^2} 
 ~\mathrm{d}s~\mathrm{d}l,
\end{align}
where $L_{\star}$ is the luminosity of the star, $n (r,s) \mathrm{d}s $ the dust number density of grains with radii 
in $[s, s+\mathrm{d}s]$ at a distance $r$, and \mbox{$l=\pm(r^2-b^2)^{1/2}$} the line of sight. 
$Q_{\text{sca}}$ and $S_{11}(\phi)$ are the scattering efficiency and the scattering phase function of the grains, respectively.
The latter can be retrieved from the $4\times4$ Mueller matrix $S_{ij}$ and 
is a function of the scattering angle $\phi=\arcsin(b/r)$.
We used Mie theory to determine $Q_{\text{sca}}$ and $S_{11}(\phi)$. 
The \texttt{ACE} simulations provided the dust number density.

To compare with the observed $V$- and $H$-band surface brightness profiles presented in \cite{Fitzgerald2007}, 
we integrated $S(b)$ over the filter transmission curves of the HST Advanced Camera for Surveys F606W 
(central wavelength $\lambda_\mathrm{c}=0.59\,\microns$, width $\Delta\lambda=0.23\,\microns$) and the Keck NIRC2 camera 
($\lambda_\mathrm{c}=1.63\,\microns$, $\Delta\lambda=0.30\,\microns$).
The disc surface brightnesses were expressed relative to the star for 
which we found 8.55\,mag in $V$-band and 4.84\,mag in $H$-band with our photosphere model.
To account for the extended halo of small particles, 
the flux was integrated up to stellocentric distances of $r=150$\,au for each line of sight. 
We also compared with the radial profiles extracted from the high-resolution HST/STIS image by \cite{Schneider2014}. 
In their Fig. 39, the surface brightness is given for both sides of the disc along with their exponentional fits.
We averaged both fits and converted the flux from mJy/arcsec$^2$ to mag/arcsec$^2$ using the zero-point magnitude flux of 3671\,Jy
and then subtracted a stellar $V$-band brightness of 8.837\,mag according to Table~1 in \cite{Schneider2014}.
The resulting $V$-band profile was found to be in good agreement with the \cite{Fitzgerald2007} data.

Figure~\ref{fig:scattered_light} depicts the modelled and observed $V$- and $H$-band profiles as well as the disc colour, $V-H$. 
All models show significant deviations from the surface brightness data in both bands. 
Although roughly reproducing the observed slopes for \mbox{$b<40$\,au}, the synthetic profiles are steeper beyond, which is the halo zone,
but get shallower with increasing stellar wind strength there.
In terms of $V-H$, the SW0 and SW50 discs appear slightly blueish (\mbox{$V-H<0$}), whereas the SW300 disc is red (\mbox{$V-H>0$}), 
which completely disagrees with the observations.
These effects are mainly due to the distribution and the amount of small grains in the disc.
Increasing the stellar wind strength pushes more small particles in the halo zone (Fig.~\ref{fig:NR_SEDs_tau}), 
and helps to enhance the scattered light flux beyond 40\,au to approach the observed level. 
However, also the blowout size increases and the whole disc becomes more and more populated by larger particles 
that are stronger scatterers at longer wavelengths. 
Hence, the disc colour $V-H$ switches to red for too strong winds, clearly conflicting with the observational data.

Next, we considered the degree of linear polarisation, given by 
\begin{align}
 P(b)=\frac{\int \int p(\phi)~ S_{11}(\phi)~ \pi s^2 Q_\mathrm{sca}~ n(r,s)~\mathrm{d}s~\mathrm{d}l}
 {\int \int S_{11}(\phi)~ \pi s^2~ Q_\mathrm{sca}~ n(r,s)~\mathrm{d}s~\mathrm{d}l},
\end{align}
where $p(\phi)$ is the linear polarisation of a single particle. 
If $P(b)$ is positive (negative), the scattered light is partially polarised perpendicularly (parallel) 
to the scattering plane, i.e. the plane containing the star, the dust grains, and the observer.

\begin{figure}[h!]
 \resizebox{\hsize}{!}{\includegraphics{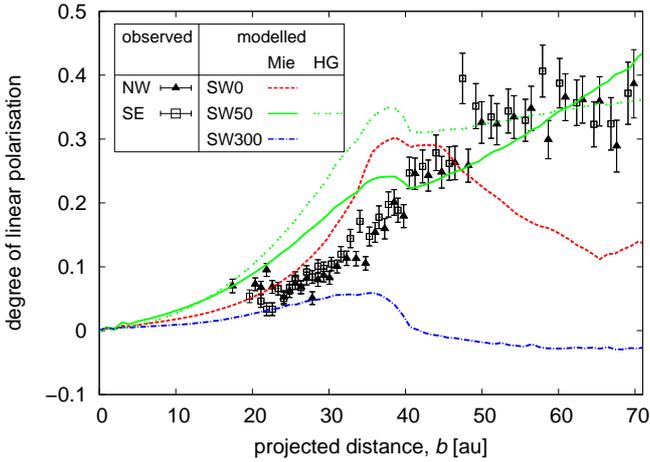}}
 \caption{Degree of polarisation as a function of projected distance. 
          Squares and triangles are measurements for the SE and NW ansae, 
          respectively \citep{Graham2007}.
          The double-dotted green line was obtained with the Henyey-Greenstein model of \cite{Graham2007}, 
          whereas the other lines are results from Mie theory.        
          }
 \label{fig:dop}
\end{figure}

As shown in Fig.~\ref{fig:dop}, the observed monotonous increase of the degree of polarisation up to $80$\,au is roughly reproduced by SW50 only.
In run SW300, the halo region is dominated by particles of $\approx0.5$\,\microns{} in size which are about one order of magnitude larger than in SW50. 
Their polarisation $p(\phi)$ strongly oscillates with the scattering angle.  
Accordingly, the polarisation integrated over a range of scattering angles along the line of sights 
tends to average to small values and $P(b)$ does not rise with distance.
For SW0, $P(b)$ increases up to the planetesimal belt as observed, but then decreases beyond. 

Although SW50 fits the measured degree of polarisation best,
the data are markedly overestimated by this model for \mbox{$b<40$\,au}.
This is caused by particles dragged from the parent belt to the inner zone.
\cite{Graham2007} and \cite{Fitzgerald2007} concluded that the inner disc region has to be relatively free of scattering grains, 
resulting in a low normal optical depth to scattering, \mbox{$\tau_\perp^\text{sca}(r)=\int \pi s^2 \,Q_{\text{sca}}\,n(r,s)\,\mathrm{d}s$}.
In $V$-band, SW50 gives \mbox{$\tau_\perp^\text{sca}(r<r_\mathrm{PB})/\tau_\perp^\text{sca}(r=r_\mathrm{PB})\geq0.1$}, 
which is at least two orders of magnitude greater than the limit found by \cite{Graham2007} and \cite{Fitzgerald2007}.
However, owing to strong stellar winds, the inward transport of a significant amount of dust is a natural outcome 
in the dynamical evolution of the outer planetesimal belt.
Even if planets exist at \mbox{$r<r_\mathrm{PB}$},
small particles come through,
as only slower, bigger particles can be efficiently scattered by planets
or get trapped in mean-motion resonances
\citep[e.g.,][]{Reidemeister2011, Shannon2015}.
It is important to note that the deviations of the model from the data may be caused 
by the assumption of spherical dust grains. 
Real particle shapes significantly deviate from spheres because the bodies underwent 
a rich accretional and collisional history.
While effective medium Mie spheres provide a good approximation to the total scattering cross sections of irregular particles,
 they cannot reproduce the scattering phase function and polarisation of such grains \citep{Shen2008, Shen2009}.
Hence, Mie theory fails in reproducing the scattered light observations as it was recently shown for HR~4796~A \citep{Milli2015, Perrin2015}.  
In a study of the HD~181327 disc, \cite{Stark2014} also found an empirical 
scattering phase function, which is not fully reproducible with Mie theory, indicating strongly forward scattering grains.
Thus, if the AU~Mic disc is populated by similar particles, our models would poorly fit 
the scattered light measurements by using Mie theory.

\begin{figure}[b!]
 \resizebox{\hsize}{!}{\includegraphics{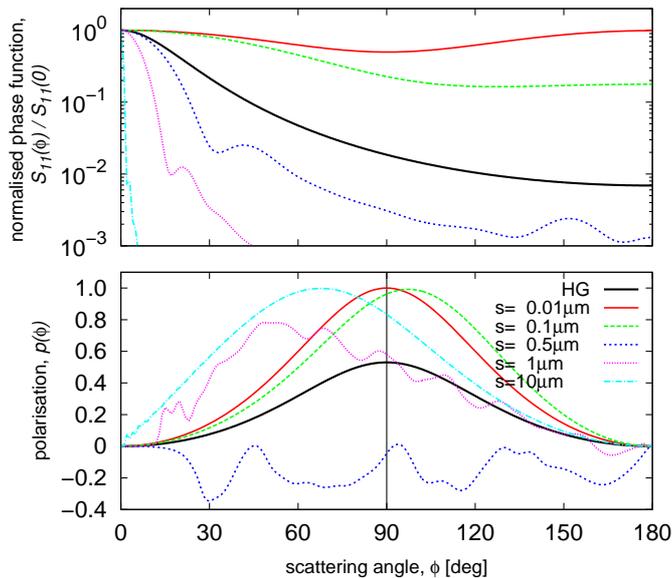}}
 \caption{Normalised scattering phase function (\emph{top}) and polarisation (\emph{bottom}) vs. scattering angle in $V$-band.
 Solid black lines have been inferred by \cite{Graham2007} via a Henyey-Greenstein model. 
 The coloured lines are for Mie spheres with radii $s$, made of material M1.
 }
 \label{fig:SPF_Pol}
\end{figure}

In an additional test, we aim at illustrating the effect when deviating from the Mie sphere dust model. 
To this end, we adapted the \cite{Graham2007} best-fit Henyey-Greenstein (HG) model for 
$S_{11}(\phi)$ and $p(\phi)$, and re-computed the $V$-band surface brightness profile and the degree of polarisation for SW50 
(double-dotted curves in Fig.~\ref{fig:scattered_light}b and Fig.~\ref{fig:dop}).
Note that $S_{11}(\phi)$ and $p(\phi)$ of the HG model are independent from grain size. 
Thus, all disc objects have the same scattering properties.
The new synthetic $V$-band profile fits better the absolute values of the surface brightness flux for \mbox{$b<40$\,au}, 
whereas the degree of polarisation rises too steeply in that range.
Reasons for this behaviour can be identified by comparing  
$S_{11}(\phi)$ and $p(\phi)$ 
predicted by Mie theory and the HG model (Fig.~\ref{fig:SPF_Pol}).
The HG phase function is distinctly broader than for Mie spheres with \mbox{$s>0.5\,\microns$}.
That is why bigger particles can scatter more light towards the observer that 
leads to the $V$-band flux enhancement for \mbox{$b<40$\,au} as shown in Fig.~\ref{fig:scattered_light}b.
Beyond 40\,au, only small particles are present and 
the HG profile resembles the one of the Mie spheres.

However, the HG model shows strong deviations from the observed degree of polarisation,
attributable to the form of the polarisation curve as a function of scattering angle (Fig.~\ref{fig:SPF_Pol}). 
The maximum of the polarisation occurs at \mbox{$\phi_\mathrm{max}=90^\circ$}, so that the main contribution to the polarised light
comes from particles with distances from the star nearly equal to their projected distances (\mbox{$r\approx b$}).
The $\phi_\mathrm{max}$ of the Mie sphere polarisation functions strongly vary with $s$.
This agrees with the trends found for irregularly-shaped particles \citep{Shen2009}.
With such shifts of $\phi_\mathrm{max}$ the polarised light at a certain $b$ mainly originates from particles at \mbox{$r>b$}.
Furthermore, irregular grains smaller than 0.1\,\microns{} tend to be weaker polarisers than Mie spheres in the same size range.
Both may help to generate a shallower increase of the degree of polarisation with projected distance.
Thus, the consideration of irregular particles in the scattered light analysis seems to be an attractive possibility 
to mitigate the deviations between the observed and modelled degree of polarisation.

 \begin{figure*}[htb!]
 \includegraphics[width=0.5\textwidth]{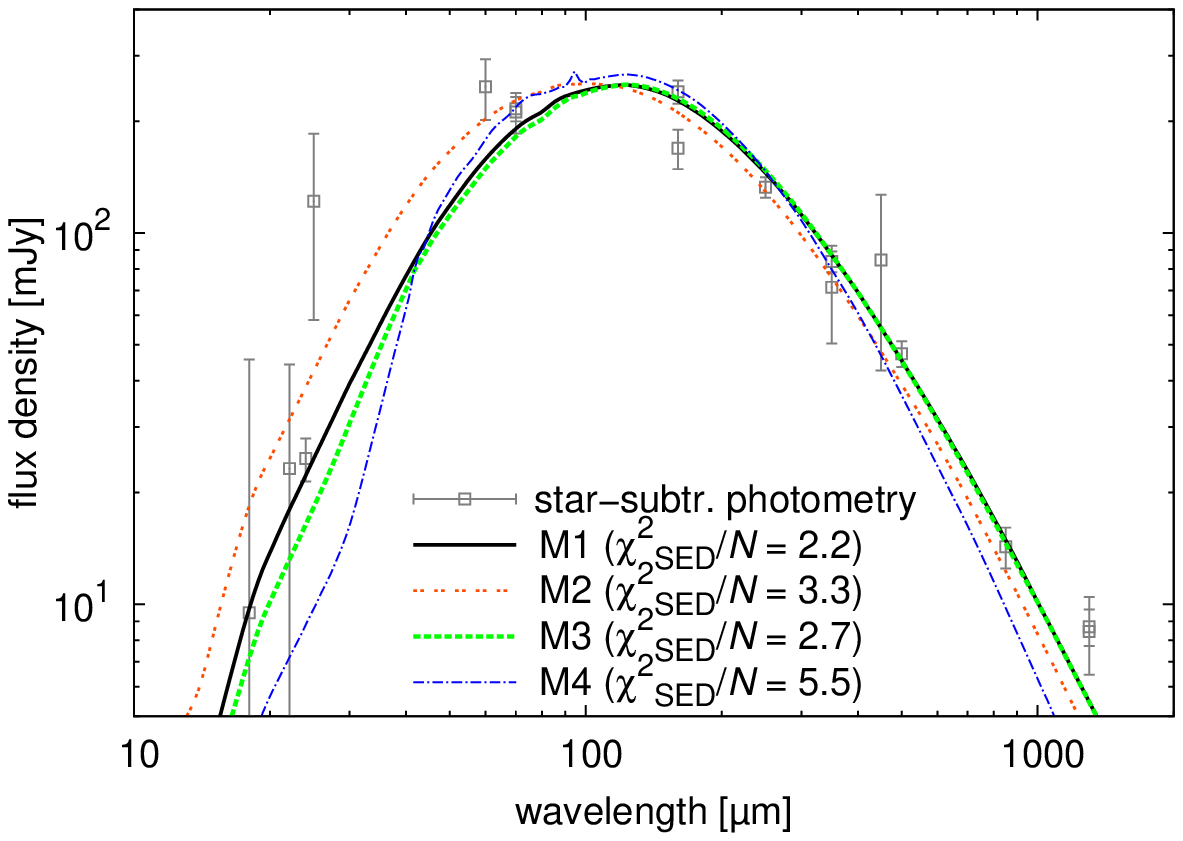}
 \includegraphics[width=0.5\textwidth]{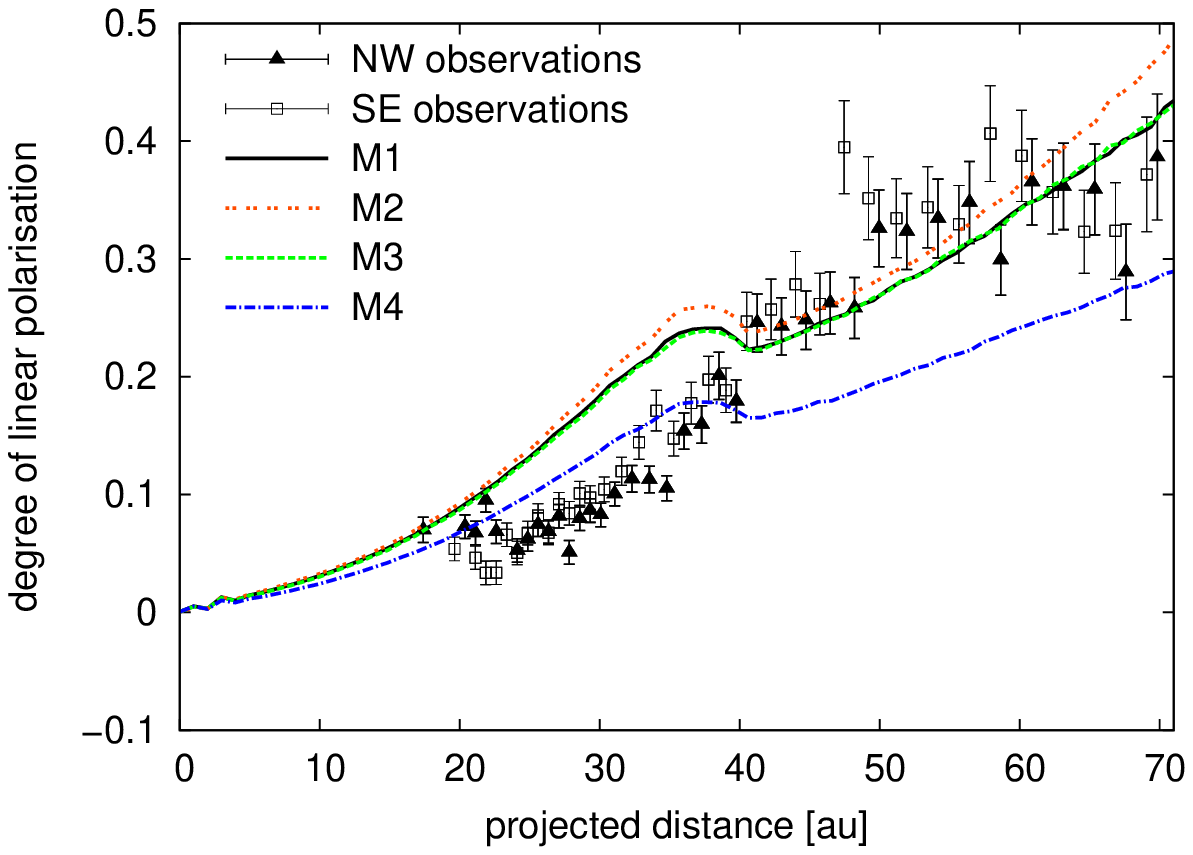}
 \caption{
 Influence of the dust composition on the SED (\emph{left}) and the degree of polarisation (\emph{right}). 
 The material names M1 -- M4 are explained in Table~\ref{tab:materials}. For better comparison of the SED models,
 $\chi^2_\mathrm{SED}/N$ values are given.
 }
 \label{fig:NR_material}
\end{figure*}

We summarise that all our models have a large amount of small grains within the parent belt in contrast to what 
was found in previous studies and none of our models provides good fits to the scattered light data.
There are two ways to interpret this result:
(1) Our models may predict the actual dust distribution in the disc well
but Mie theory is not valid to simulate the scattering properties of the grains. 
Then, our analysis show a preference towards a scattered light model where the grains
are weaker polarisers than Mie spheres with polarisation maxima 
at scattering angles different from $90^\circ$.
(2) Mie theory is an appropriate technique for the scattered light analysis but 
there are shortcomings with respect to the dust density derived in the collisional modelling.
The truth may also be a combination of (1) and (2). 
A consideration of point (1) goes along with the simulation of the emission from irregular particles which is
beyond the scope of this work. 
Thus in the following, we consider only point (2) in more detail. 
Since the scenario SW50 is found to be in best agreement with the data amongst the three stellar wind strengths tested, 
we consider SW50 a reference model, which is the starting point for further individual parameter 
variations influencing the dust distribution.

\subsection{Material composition}
\label{sec:material_composition}

The chemical composition of the AU~Mic disc  might be similar to discs around other BPMG stars because 
they have formed in the same molecular cloud from the same material and are coeval.
However, the disc objects are heavily processed during the debris disc and preceding protoplanetary disc phases.
Thus, their composition may depend on the mass and luminosity of the host star.
Circumstellar dust compositions for some BMPG members have already been analysed previously.
\cite{Smith2009} and \cite{Churcher2011} considered grains with a silicate core and a mantle of organic refractories 
for SED modelling in the $\eta$~Tel and HD~191089 systems. 
They pointed out that particles with small silicate cores and a porosity of 20\% ($\eta$~Tel) and 60\% (HD~191089)
provide good fits to the observed excesses. 
A similar model that incorporates porous core-mantle silicate grains, additionally covered by ice in the outer region of the disc, 
also reproduces the $\beta$~Pic observations \citep{Pantin1997, Li1998, Augereau2001}.
\cite{Lebreton2012} tested various chemical compositions for the debris around HD~181327. 
They found the SED to be mostly consistent with grains consisting 
of amorphous silicate and carbonaceous material with a dominant fraction of ice. 
In addition, the particles have to have a porosity of $65\%$.

The material composition affects the dust distribution and the SED by 
changing $\beta$  and the dust temperature, $T_\mathrm{d}$ 
(e.g., \citeauthor{Kirchschlager2013}~\citeyear{Kirchschlager2013} 
for the influence of porosity on $\beta$ and $T_\mathrm{d}$).
We now depart from the mixture M1, used in our reference run SW50, to assess whether other materials 
can be viable for the AU~Mic system.
We repeated the simulation with the compositions M2, M3, and M4  explained in Table~\ref{tab:materials}, 
and investigated modifications to observables  (Fig.~\ref{fig:NR_material}).

The material M2 represents compact particles, composed of  silicate and carbon.
For these, we found a shift of the  SED towards shorter wavelengths that 
deteriorates the fit in the submillimetre to radio range.
In M3 and M4, we included water ice additionally. 
All constituents are present in equal volume fractions in M3, whereas  
the porosity and ice fraction are increased in M4.
M3 is in fairly good agreement with the observed photometry, 
only tending to underestimate the near-IR excess. 
The icier and more porous M4 produces an SED whose peak position
remains rather unchanged but increases and decreases more steeply. 
This narrower SED distinctly disagrees with the near-IR and submillimetre measurements.

No large differences in terms of the degree of polarisation are noticeable for M1, M2, and M3, 
whereas M4 markedly underestimates the data.
The latter gives evidence that the material must not be too icy and porous.
Note that modelling of the degree of polarisation is possibly 
affected by limitations of the Mie theory  (see Sect.~\ref{sec:scattered_light}) 
that does not allow us to find strong constraints for the upper limit of 
the ice and vacuum content.

We noticed little changes in the radial profile at 1.3\,mm for all materials considered.
This is caused by two characteristics of large grains, dominating the long-wavelength emission.
First, large grains are weakly affected by stellar radiation and stellar winds 
and their spatial distribution is equal for different materials. 
Second, the dust temperature approaches the blackbody temperature with increasing particle size
at a given distance \citep[e.g., Fig.~5 of][]{Pawellek2014}.
Therefore, $T_\mathrm{d}$ is nearly independent of the chemical composition  for large grain sizes.

Altogether, the SED and the degree of polarisation seem to be consistent with materials 
having a small fraction of ice and a moderate porosity.

\subsection{Dynamical planetesimal excitation}
\label{sec:excitation}

\begin{figure}[b!]
\vspace*{-0.5cm}
 \centering
 \resizebox{\hsize}{!}{\includegraphics{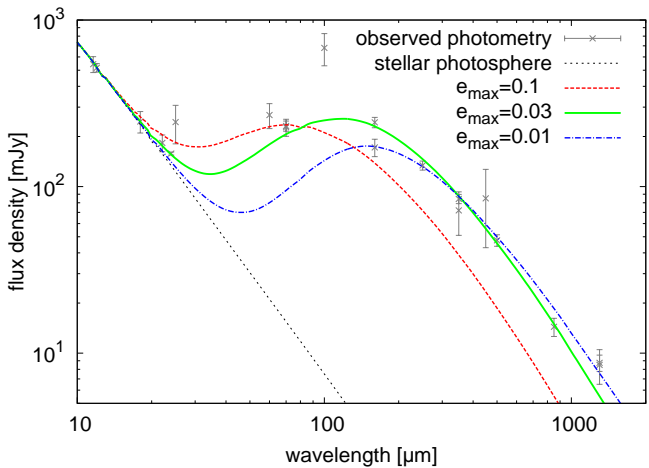}}
 \caption{SED models for different values of the 
 maximum eccentricity, $e_\mathrm{max}$, of the planetesimals orbits.
 }
 \label{fig:NR_SEDs}
\end{figure}
\begin{figure*}[b!]
 \resizebox{\hsize}{!}{\includegraphics{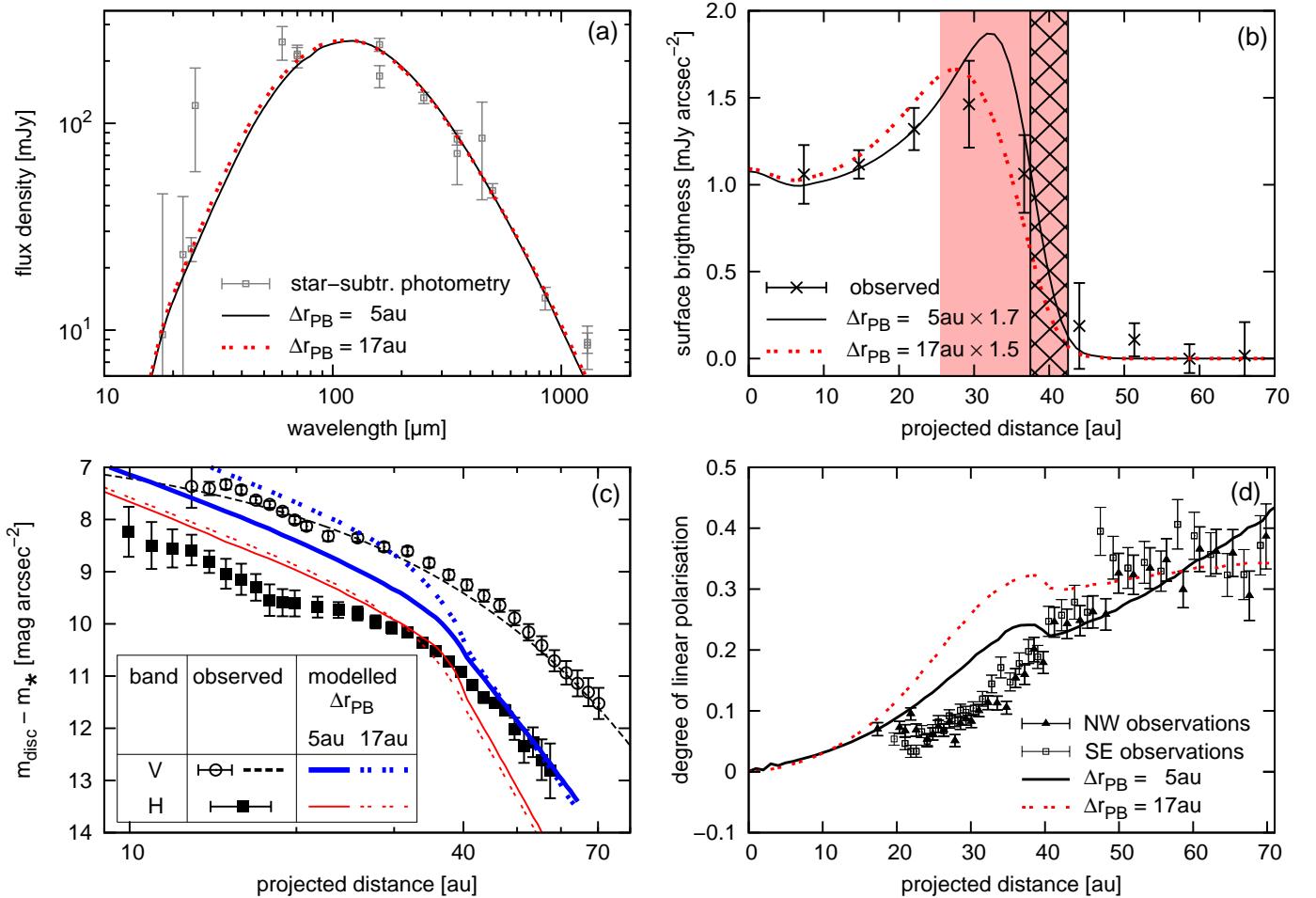}}
 \caption{
  Influence of the planetesimal belt width, $\Delta r_\mathrm{PB}$, 
 on \emph{(a)} the SED, \emph{(b)} the 1.3~mm radial profile,
 \emph{(c)} the $V$- and $H$-band profiles,
 and \emph{(d)} the degree of polarisation. 
 The filled patterns in panel \emph{(b)} depict the location and the radial width 
 of the planetesimal belts (black-hatched 5\,au, red-shaded 17\,au). 
 The profiles were vertically scaled by 1.7 ($\Delta r_\mathrm{PB}=5$\,au) and 1.5 ($\Delta r_\mathrm{PB}=17$\,au) 
 to compensate the modelling imperfections mentioned in Sect.~\ref{sec:thermal_emission}.
  }
  \label{fig:broader_birth_ring}
\end{figure*}%

The maximum eccentricity of the planetesimals, $e_\mathrm{max}$, represents the disc's dynamical excitation
since this parameter affects the impact velocities of the objects, and therefore, the 
amount of small grains produced in collisions. 
In addition to the reference model with \mbox{$e_\text{max}=0.03$}, we consider
\mbox{$e_\text{max}=0.01$}, a low value corresponding to the expected maximum level of pre-stirred discs \citep[][and references therein]{Matthews2014},
and \mbox{$e_\text{max}=0.1$}, a high value representing the dynamical excitation in the Edgeworth-Kuiper belt \citep[e.g.,][]{Vitense2010}.
The resulting SEDs are shown in Fig.~\ref{fig:NR_SEDs}.

Increasing $e_\text{max}$ favours the production of small particles and shifts
the SED maximum to shorter wavelengths.
Thus, the model with \mbox{$e_\text{max}=0.1$} well agrees with the excess data at \mbox{$\lambda<70\,\microns$},
but completely fails to reproduce the ones at longer wavelengths.
In contrast, \mbox{$e_\text{max}=0.01$} provides a good match to the 
photometry beyond 100\,\microns{}, but strongly deviates
from the measurements at \mbox{$\lambda\leq70\,\microns{}$}.
Improvements may be possible by additional variations of other parameters. 
For instance, shifting the planetesimal belt inwards (outwards) 
increases (decreases) the dust temperature and 
helps to find a better SED fit for \mbox{$e_\mathrm{max}=0.01$} (\mbox{$e_\mathrm{max}=0.1$}).
However, any significant $r_\mathrm{PB}$-shift would result in stronger deviations from 
the ALMA 1.3\,mm profile, which constrains the location of a narrow planetesimal belt at around 40\,au.
Another possibility would be broadening the SED towards shorter wavelengths for \mbox{$e_\mathrm{max} = 0.01$}. 
This can be achieved by stronger stellar winds as depicted in Fig.~\ref{fig:NR_SEDs_tau}.
However, stronger winds also lead to an increase of the blowout size, and therefore, it is more difficult to 
reproduce the observed blue disc colour (Sect.~\ref{sec:scattered_light}). 
Furthermore, other dust materials may have a positive effect on the SEDs.
None of the materials investigated in Sect.~\ref{sec:material_composition} strengthens the submillimetre flux density, 
as required to improve the SED for \mbox{$e_\mathrm{max}=0.1$}. 
Only one composition -- the non-porous M2 material -- predicts a stronger mid-IR emission 
from which the low-$e_\mathrm{max}$ model would benefit, 
but this material agrees less with the scattered light measurements.
In summary, the reference run with \mbox{$e_\text{max}=0.03$} matches the entire SED the best. 
Although values higher or lower than 0.03 seem to be unlikely,  
they cannot be ruled out completely if combined with other parameters, which was not explored here.

An observational constraint on the dynamical excitation
can be put by the vertical thickness of the disc through the 
equilibrium condition $i=e/2$, where $i$ and $e$ are the orbital inclinations and eccentricities of the disc particles.
Thus, measuring the disc's opening angle at long wavelengths hints at the maximum eccentricity of the planetesimals.
However, the AU~Mic disc is vertically unresolved in the SMA and ALMA images \citep{Wilner2012,MacGregor2013}, which 
gives only upper limits of the disc extent in this direction. 
We compared synthetic vertical profiles of our \mbox{$e_\mathrm{max}=0.1$} model   
with the ones extracted from the ALMA image.
The modelled profiles are marginally consistent with the ALMA data, constraining 
the disc's semi-opening angle to \mbox{$i_\mathrm{max}\lesssim e_\mathrm{max}/2 = 0.05$}.
Interestingly, \cite{Krist2005} and \cite{Metchev2005} derived \mbox{$i_\mathrm{max}=0.04-0.07$} 
from visible and near-IR images.
Reading off the vertical scale height from the \cite{Schneider2014} high resolution STIS image
gives \mbox{$i_\mathrm{max}=0.03$}.
Thus, the observations yield a maximum eccentricity \mbox{$e_\mathrm{max}=0.06-0.14$} that is excluded
by our modelling.
Resolving this contradiction may be possible if the vertical disc thickness seen in short-wavelength images does not 
necessarily points towards the dynamical state of the planetesimals.
\cite{Thebault2009} proposed a possibility for the AU~Mic disc to develop a 
vertical thickness close to the observed value in the visible and near-IR although the disc is dynamically cold. 
In that model, the smallest grains have high in-plane velocities
due to radiation and stellar wind pressure, and these velocities are partially converted into vertical ones 
in collisions with other particles.
As a consequence, the disc becomes naturally thicker even without the gravitational perturbation 
of large, massive embedded bodies. In equilibrium, the disc has a large 
vertical dispersion of the smallest grains, whereas large particles remain close to the midplane.
Thus, a low $e_\mathrm{max}$, as preferred in our collisional 
modelling, can be consistent with the observed vertical disc thicknesses at short wavelengths.

\begin{figure*}[t!]
 \includegraphics[width=0.5\textwidth]{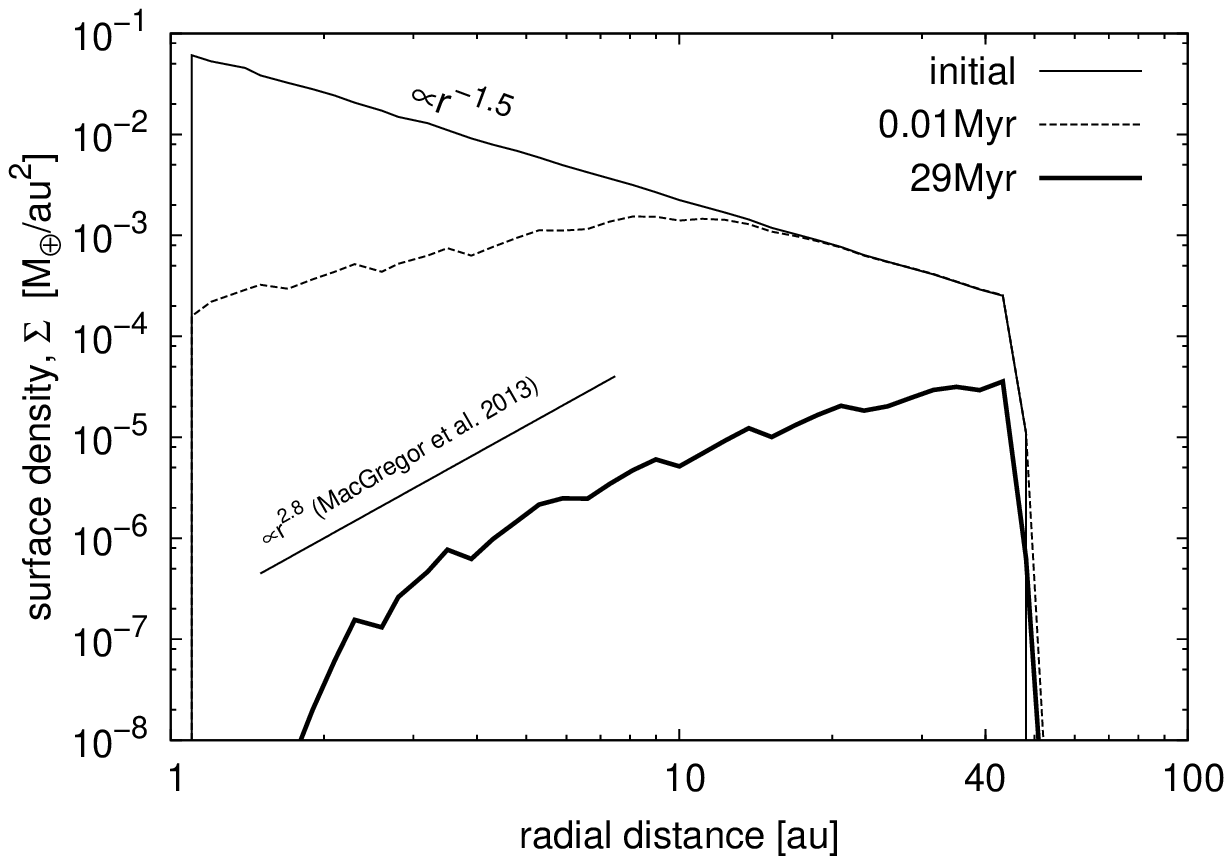}
 \includegraphics[width=0.5\textwidth]{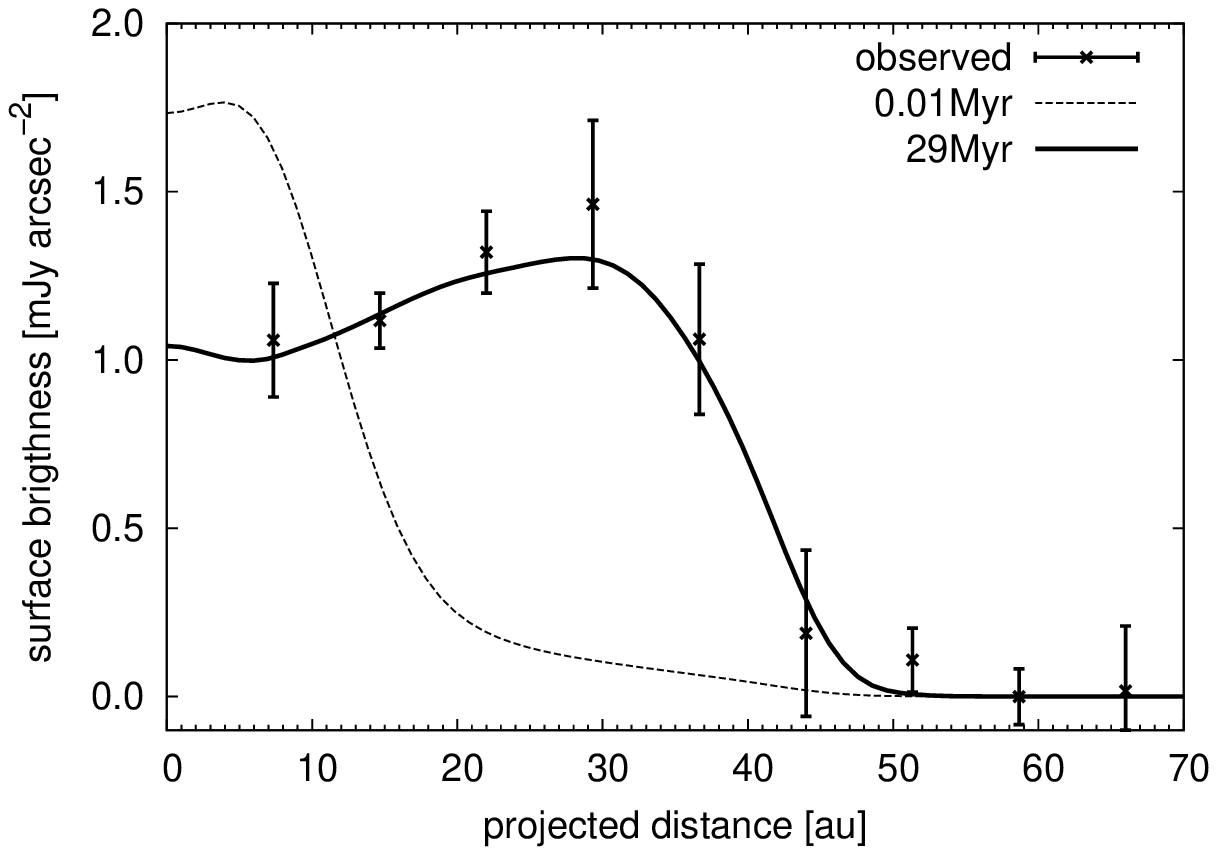}
 \caption{\emph{Left}: time evolution of the surface density in an extended planetesimal disc (\mbox{$\Delta r_\mathrm{PB}\approx45$\,au}). 
 The wavy structures of the $\Sigma$-curves are numerical artefacts.
 For the sake of comparison, the straight line shows the $\Sigma$-slope derived by  \cite{MacGregor2013}.  
 \emph{Right:} time evolution of the 1.3\,mm profile along with the ALMA data. 
 Note that the 0.01\,Myr surface brightness profile was scaled by a factor of $0.05$ for better visibility.
  }
 \label{fig:BD_sigma}
\end{figure*}

\subsection{Planetesimal belt width}
\label{sec:belt_width}
\subsubsection{17\,au-wide belt}
\label{sec:17au-wide_belt}

As noted in Sect.~\ref{sec:thermal_emission}, 
the outer edge of the planetesimal belt is well constrained by the ALMA profile. 
However, the edge-on orientation of the disc and the limited 
information on the innermost regions affected by the bright central source
of unknown nature make the position of the inner edge of the disc, and so the disc extent, quite uncertain.
In another test, we increased the planetesimal belt 
width from \mbox{$\Delta r_\mathrm{PB}=5$\,au} (reference model) to \mbox{$\Delta r_\mathrm{PB}=17$\,au}
by shifting the inner edge to $25.5$\,au while keeping the outer edge at $42.5$\,au. 
The initial surface density was assumed to be constant over the belt width.
Figure~\ref{fig:broader_birth_ring} compares the 17\,au-wide belt scenario
with the reference model.

The broader planetesimal belt has nearly the same SED as in the reference run and 
does not increase the warm emission although
the dust production zone is more extended towards the star.
The 1.3\,mm profile of the 17\,au-wide belt peaks closer to the star than
for the reference model.
Thus, it tends to underestimate the observed surface brightness farther out.
This shortcoming does not hint at an implausibility of a broader planetesimal belt model  since it 
 can easily be compensated by shifting the outer edge from $42.5$\,au outwards.
Although the broader belt scenario matches better the $V$-band profile than the reference run does, 
the degree of polarisation shows stronger deviations which may
 be attributed to the light scattering model (Sect.~\ref{sec:scattered_light}).

\subsubsection{44\,au-wide belt}
\label{sec:40au-wide_belt}

We now address the possibility that the AU~Mic disc was populated by planetesimals over a wide range of distances.
In such a disc, the innermost planetesimals have the shortest collisional lifetimes and 
are depleted within a few Myr, resulting in 
a surface density increasing outwards \citep{Kennedy&Wyatt2010}.
Due to the inside-out erosion of objects, an extended planetesimal disc may look like a narrow belt in long-wavelength
observations, even if there are no planets within the disc that have swept up the inner region.
In their study of the collisional evolution of a fiducial, 100\,au-wide disc,
\cite{Thebault2007} confirmed that the surface brightness peaks at (sub)millimetre wavelengths close 
to the outer edge of the planetesimal distribution after several Myr.
Thus, the extended-disc scenario is an auspicious alternative for explaining 
the rising ALMA profile up to projected distances of $\approx$\,30\,au in the AU~Mic system. 
The significant fall beyond 30\,au may mark the point beyond which planetesimals 
have not formed at all or where the disc was truncated.

To assess the feasibility of this scenario, we started an 
\texttt{ACE} simulation with a broad radial distribution of planetesimals from 1 to 45\,au. 
Here we already accounted for a slight shift of the outer edge from the reference value \mbox{$42.5$\,au} to 45\,au, as suggested 
in Sect.~\ref{sec:17au-wide_belt}.
The run was initialised with a solid surface density given 
by the Minimum Mass Solar Nebula (MMSN) model, \mbox{$\Sigma \propto r^{-1.5}$}.
We considered bodies up to 100\,m in radius and 
assumed pre-stirred planetesimals with \mbox{$e_\mathrm{max}=0.03$} (Sect.~\ref{sec:excitation}).

Figure~\ref{fig:BD_sigma} depicts the evolution of the surface density and the 1.3\,mm profile. 
The surface density quickly dropped close to the star due to particle erosion and 
the initially negative slope ($-1.5$) switched to a positive slope in a few ten thousand years.
Note that the particles' collisional lifetimes were shorter than their orbital periods at the beginning of 
the simulation due to the large particle density at that phase. 
Thus, two consecutive collisions of one object are correlated 
and the orbit averaging method, implemented in \texttt{ACE}, is a crude approximation to 
determine the phase space density. 
Therefore, the results for the first one hundred thousand years of simulation time should be treated with caution.

Owing to the progressive inside-out depletion of the planetesimals, the maximum of the 1.3\,mm profile
moved towards larger distances as time elapses.
After 3\,Myr, we noticed that the profile peak reached a maximum distance of about 30\,au, which is in good agreement with the ALMA data.
The absolute flux level of the observed dust emission was reproduced after 29\,Myr.
The final run fits the SED, the scattered light profiles, and the degree of polarisation with nearly the same fidelity 
as the 17\,au-wide belt model (Sect.~\ref{sec:17au-wide_belt}).

Note that the collisional evolution of the disc was artificially shortened as the 
largest bodies in the disc had been 100\,m in size.
If the disc is contains larger bodies, the inside-out planetesimal depletion is slower and the disc 
needs more time to reach the collisional equilibrium \citep{Loehne2008}. 
As a result, it takes longer than 3\,Myr to reach the observed shape of the 1.3\,mm profile.
This way, scaling the model to the proper age of the system is possible.
However, we refrained from further simulations, 
because of the numerical complexity of the extended planetesimal 
disc simulation\footnote{It took about 150\,CPU days to evolve the disc over 29\,Myr.}.
The run presented does illustrate the feasibility of an extended planetesimal disc scenario sufficiently well.


\section{Unresolved central emission}
\label{sec:central_emission}
In Sect.~\ref{sec:outer_disc}, we did not care about the nature of the inner unresolved component and
assumed its contribution to the flux densities at all wavelengths to be negligible.
However, this may not be true if the unresolved central emission originates from an 
inner dust ring. The dust there would be much warmer than in the outer disc
and would emit significantly at short wavelengths, while adding little flux in the mm range.
We now consider the possibility of an additional inner planetesimal ring in the AU~Mic system.
If such a ring could exist, we should also investigate its consequences for the modelling of the resolved outer disc, discussed 
in Sect.~\ref{sec:outer_disc}.

\begin{figure}[h!]
 \resizebox{\hsize}{!}{\includegraphics{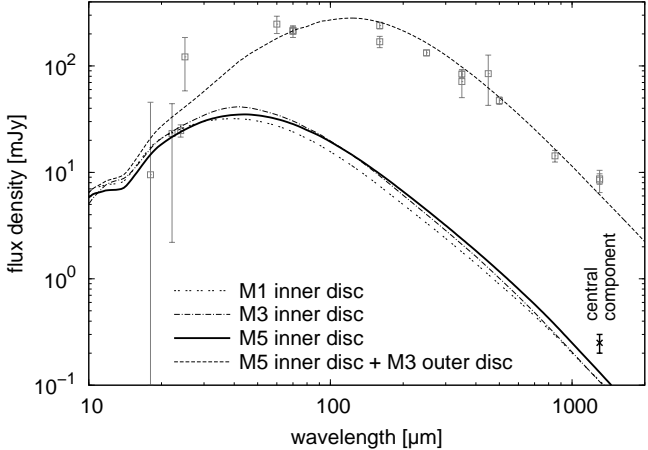}}
 \caption{
 SEDs of the inner planetesimal rings centred at \mbox{$r_\mathrm{PB,in}=3$\,au} (width \mbox{$\Delta r_\mathrm{PB,in} = 0.4$\,au}) for
 the compositions M1, M3, and M5 (Table~\ref{tab:materials}).
 Squares illustrate the photosphere subtracted photometry. 
 The black cross in the right lower corner shows the emission of the unresolved star subtracted central component at 1.3\,mm
 with $1\sigma$ uncertainty. 
 The dashed line represents the M3 outer disc in Fig.~\ref{fig:NR_material} in 
 combination with the inner M5 disc.
 \label{fig:inner_ring}
 }
\end{figure}

A putative inner dust ring has to be 6 times brighter 
than the stellar photosphere at 1.3\,mm \citep[][and Sect.~\ref{sec:ALMA_observations} of this paper]{MacGregor2013}.
Simultaneously, the ring has to emit weakly in the mid-IR in order not to contradict  
the measured excess there.
Since an ALMA resolution of about 0.6'' ($\approx6$\,au) was achieved, the ring 
should be located at distances $\lesssim$\,3\,au to be compatible with remaining unresolved.
We placed an inner planetesimal ring at \mbox{$r_\mathrm{PB,in}=3$\,au}, 
having a full width \mbox{$\Delta r_\mathrm{PB,in}=0.4$\,au}, and assumed $e_\mathrm{max}=0.03$ and 50 times the
solar wind strength. 
We made sure that the ring emission did not exceed the 
excess shortward of 24\,\microns{} and then investigated its emission at 1.3\,mm.
Figure~\ref{fig:inner_ring} shows the results.

First, we assumed two inner rings with M1 and M3 -- materials we found 
suitable for the outer component (Sect.~\ref{sec:material_composition}).
Consistency with the short-wavelength excess was achieved with a maximum dust mass 
of \mbox{$\approx5\times10^{-6}\,M_\oplus$}. 
The rings are not bright enough to completely reproduce the central emission.
Although M3 enhances the SED maximum, the long-wavelength slope is steeper and 
levels to nearly the same flux density at 1.3\,mm as for M1.
In another attempt, we considered highly porous grains (M5).
This is motivated by the possibility that icy grains are produced beyond the ice 
line and then drift towards the star by drag forces.
Accordingly, the particles get heated and their sublimated ice fractions are replaced by  
vacuum inclusions, which increase the porosity.
Assuming an ice sublimation temperature \mbox{$T_\mathrm{subl}=100$\,K} 
\citep[e.g.,][]{Kobayashi2008} yields an iceline at  
\mbox{$r_\mathrm{ice}=(L_\star/L_\odot)^{0.5}\,(T_\mathrm{subl}/277\,\kelvin)^{-2}\,\au\approx2\,\au$}, which 
is close to $r_\mathrm{PB,in}$.
The M5 material emits slightly more than M1 and M3 in the mm regime, but reproduces only about
60\% of the observed central flux density. 
This constrains the maximum fraction of the central emission that can be explained by pure dust emission.
Thus, the central emission must originate only or in parts from another source. 
Note that the star itself can contribute to the central emission, for instance
due to an active stellar corona \citep{Cranmer2013}. 
Furthermore, stars may emit more flux at long wavelengths than predicted by the
purely photospheric models for their spectral class. 
ALMA observations of the G- and K-type stars 
$\alpha$~Cen A and B show a flux enhancement by 20--30\% in band 7 and
almost doubling of the flux in band 3 \citep{Liseau2015}. 
Provided such a trend continues to an M dwarf like AU~Mic, we do expect
a significant stellar contribution to the observed central emission.

To get the total SED, the inner ring should be combined with an appropriate model for the outer disc.
In order to avoid contradictions with the data, the outer disc has to emit less flux in the mid-IR 
where the SED of the inner ring peaks.
This is well achieved by the M3 outer disc (Sect.~\ref{sec:material_composition}), and indeed, 
in combination with the M5 inner ring the overall SED is well reproduced (Fig.~\ref{fig:inner_ring}, dashed line).
Furthermore, a combination of the icy material in the outer ring and 
iceless, highly porous material in the inner ring is reasonable, as explained above.
We adopted the outer disc SED from the model described in Sect.~\ref{sec:material_composition} without 
any adjustments (e.g., vertical scaling) that one possibly expects due to the presence of the inner ring. 
Thus, even if the inner emission stems from dust, it seems to barely affect the outer disc modelling.
The $\chi^2_\mathrm{SED}/N$ increases from 2.7 for the M3 outer disc to 3.3 for the combined M5 inner and M3 outer disc.
The larger $\chi^2_\mathrm{SED}/N$ is mostly due to the deviation of the inner+outer disc model
from the \emph{Spitzer}/MIPS 24\,\microns{} point.
However, the overestimated flux density at 24\,\microns{} may be compensated by a modification of the dust composition. 
As seen in Fig.~\ref{fig:NR_material}, adding ice reduces the flux in the mid-IR,
while leaving it nearly unchanged at longer wavelengths.
This promises improvements if a material icier than M3 for the outer disc is used. 
We highlight the small shift of the blowout size when going from iceless to icy grains (Tab.~\ref{tab:materials}). 
Accordingly, the abundance of small grains will not change significantly, still reproducing the scattered light data (Sect.~\ref{sec:scattered_light}).

We stress that our two-component model does not account for the collisional interaction between inner and outer component
which is mainly caused by two effects: 
(i) collisions of inner disc particles with inwards dragged grains from the outer disc 
and (ii) collisions between unbound particles produced in the inner ring with outer disc particles. 
By comparing the normal optical thicknesses of inner and outer component, including bound and unbound grains,
we found
\mbox{$\tau_{\perp,\mathrm{out}}/\tau_{\perp,\mathrm{in}} \sim 0.1$} at 3\,au and \mbox{$\tau_{\perp,\mathrm{in}}/\tau_{\perp,\mathrm{out}} \sim 0.001$}
at 40\,au.
Thus, collisional interaction between both is weak, justifying our modelling strategy.

\begin{table*}[t!]
\caption{Power-law fitting results.}
\label{tab_fits_sand}
\centering
\begin{tabular}{cccccc}
\hline\hline
  Parameter                      & Range explored   & \# values & Distr. & Best-fit [$3\sigma$] & + inner \\
  \hline
   $R_\mathrm{in}$ [au]          & 2 -- 60      & 187       & temp   & 3.1* [2.0* -- 24.8]  & 14.5  \\
   $R_\mathrm{out}$ [au]         & 3 -- 100     & 155       & temp   & 41.9 [37.3 -- 45.5]  & 41.9  \\
   $\alpha$                      & -5 -- 5      & 101       & lin    & 2.8 [1.8 -- 4.8]     & 3.5   \\
  \hline
   $s_\mathrm{min}$ $[\microns]$ & 0.2 -- 20      & 494       & log    & 0.2* [0.2* -- 0.5]   & 1.8   \\
   $s_\mathrm{max}$ $[\microns]$ & 2000           & 1         & fixed  & \dots                & \dots \\
   $\gamma$                      & -5 -- -3     & 21        & lin    & -3.3 [-3.4 -- -3.3]  & -3.3   \\
  \hline
   $\theta$                      & 90\degr          & 1         & fixed  & \dots                & \dots \\
   $M_\mathrm{d}$ [$M_\oplus$]   & \dots            & \dots     & cont   & $4.7\times10^{-3}$   & $5.7\times10^{-3}$ \\
  \hline
\end{tabular}
\tablefoot{The distributions of the values considered in the parameter space are: \emph{temp} -- by equal steps in temperature 
of the grain size with the steepest radial temperature gradient, \emph{lin} -- linear, \emph{log} -- linear in the logarithm of the parameter, 
\emph{fixed} -- fixed value (no distribution at all), \emph{cont} -- continuous (scaling of the disc mass to minimise the $\chi^2$ for given values 
of all other parameters). Values marked with * are at (or very close to) the boundaries of the parameter space explored and cannot 
be considered as reliable. The real values may lie outside the parameter space explored. 
The column ``+ inner'' shows best-fit values of the outer disc after subtracting a model of the inner, unresolved component, assuming it to be a 
debris disc. These values illustrate the uncertainties of the model parameters due to the unknown nature 
and uncertain parameters of this component.
The mass values given are for grains up to $s=1$\,mm.
}
\end{table*}


\section{Comparison with parametric modelling}
\label{sec:power-law_model}

Collisional modelling allows us to explore only a small number of parameter combinations.
Although the models found reproduce the data satisfactorily, they are not unique, and better solutions may exist.
In order to find an independent confirmation of the collisional modelling results, we additionally analysed the data
with a multidimensional fitting approach. 
In that approach, the dust number density was assumed to be a combination of two independent power laws, 
representing the size and radial distribution of the grains.
This model does not consider the physical mechanisms of the dust production
and evolution.
However, it is well suited for an exploration of a wide parameter space to find the most likely solutions
that we compared with the collisional modelling results.

In the power-law model, the surface density follows $r^\alpha$ between the inner and outer cut-off radii, 
$R_\mathrm{in}$ and $R_\mathrm{out}$, while the grain size distribution follows $s^\gamma$ between the lower 
and upper cut-off grain sizes, $s_\mathrm{min}$ and $s_\mathrm{max}$. 
The influence of $s_\mathrm{max}$ is negligible for reasonably steep grain size distributions with $\gamma < -3$. 
We fixed $s_\mathrm{max}$ to 2\,mm.
The dust mass $M_\mathrm{d}$ is an additional free parameter.  
The disc inclination was fixed to $\theta=90$\degr.
For the dust composition, we assumed compact, spherical grains composed of pure 
astronomical silicate \citep{Draine2003b}. 
This makes a total of eight free parameters, $R_\mathrm{in}$, $R_\mathrm{out}$, $\alpha$, $s_\mathrm{min}$, $\gamma$, $M_\mathrm{d}$, 
the dust composition (sampled by only one value, namely 100\% astronomical silicate), and the vertical scaling of the ALMA profile.
For the fitting of the SED and the ALMA profile, we used the \texttt{SAnD} code \citep{Ertel2012, Loehne2012} which was originally developed for 
fitting simultaneously SED data and spatially resolved \textit{Herschel} data as part of 
the \textit{Herschel}/DUNES modelling tool box. 
Note that we did not include the scattered light data for this power-law fitting.

In a first approach, we fitted the model to the data ignoring the contribution of the central component. 
The best-fit model (Table~\ref{tab_fits_sand}) reveals features we already found, at least qualitatively, 
by collisional modelling: 
an outward-increasing surface density with an index close to $2.8$, a well constrained outer disc radius 
at about 40\,au, a small lower grain size $<$\,1\,\microns{} consistent with a moderate level of stellar winds,
and a relatively shallow slope of the grain size distribution, $\gamma=-3.3$, indicating that small particles are more affected 
by transport than collisions.
Furthermore, the results are widely consistent with earlier studies of 
the disc \citep{Augereau&Beust2006, MacGregor2013}.

In a second approach, we assumed the inner component to be an additional ring of dust. 
The ring was placed at 2\,au from the star with an extent of 0.2\,au and a constant surface density, 
which is just compatible with being mostly unresolved by the ALMA observations.
The lower grain size, $s_\mathrm{min,in}$, and 
the exponent, $\gamma_\mathrm{in}$, of the size distribution for this inner component were explored by hand.
The model was scaled to the flux density of the central component seen in the ALMA data. 
We found a very large lower grain size, $s_\mathrm{min,in}=5\,\microns$, 
as well as a very flat grain size distribution, $\gamma_\mathrm{in} = -3$, 
to be necessary in order to keep the dust cold enough (the grains large enough) 
not to produce too much excess at mid- to far-IR wavelengths.
It is important to note that this model serves to estimate the potential contribution 
of the unknown inner component, which is probably not a debris disc, to the SED.
However, the fitting result, even the large lower grain size, might be consistent
with a normal debris disc. 
A large $s_\mathrm{min,in}$ is naturally expected if the grains do not experience a 
radiation pressure blowout limit as is the case for AU~Mic.
Then, the dominant or critical grain size, $s_\mathrm{c}$, is the one where collisional and transport lifetimes are equal, 
which can be reached at much larger sizes.
Including the effect of stellar wind drag in Eq.~(6) of \cite{Kuchner&Stark2010}, and 
assuming an optical depth of $10^{-3}$ at 3\,au in accordance with our M5 inner ring model, 
yields a critical size \mbox{$s_\mathrm{c}\approx3$\,\microns}, which is not far 
from \mbox{$s_\mathrm{min,in}=5\,\microns$}.

In a final step, we subtracted the contribution of the inner ring from the SED data and re-fitted the remaining 
flux by the two-power-law model used for our first approach 
(Table~\ref{tab_fits_sand}, last column). 
Comparing with the previous results illustrates the uncertainties of our first approach.
The values are almost all within the 3$\sigma$-confidence interval of the first fit. 
This demonstrates that the inner component marginally 
influences the modelling of the resolved outer disc, as already found by collisional modelling.


\section{Conclusions}
\label{sec:conclusions}

We have performed in-depth collisional modelling of the AU~Mic debris disc for the first time. 
A wealth of observational data have been considered, including the densely sampled  
spectral energy distribution, 
the ALMA 1.3\,mm thermal emission profile,
the scattered light profiles in $V$- and $H$-band, 
and the degree of scattered light polarisation 
as a function of projected distance in the disc midplane.
Our study presents the first attempt to reproduce 
scattered light observations of a debris disc with collisional modelling.

Our models provide generic radial and size distributions of the particles in the whole disc.
However, the disc also possesses several substructures and asymmetries like bumps and warps \citep{Liu2004, Schneider2014} 
that may be caused by the perturbing effects of so far undetected planets.
As our modelling technique is limited to axisymmetric discs,
the formation of such disc properties was not considered in this work.

Most of the data can be reproduced with a narrow belt of planetesimals
centred around 40\,au with strong inward transport of dust by stellar winds, 
according to the birth-ring scenario of \cite{Strubbe&Chiang2006}.
For modelling the ALMA 1.3\,mm profile, a significant scaling factor $\lesssim$\,2 was necessary.
This mainly hints at the inability of our simulated dust distributions to fully reproduce the 
observed ALMA flux density.
Additionally, the synthetic dust material used may underestimate the dust emissivity at mm wavelengths.
We find a clear preference for a belt with a low dynamical excitation 
where the planetesimals orbits have a maximum eccentricity of $0.03$.
Our modelling allows us to make inferences on the stellar mass loss rate $\dot{M}_\star$, 
which is a measure for the stellar wind strength. 
We can exclude extreme values of hundreds of the solar mass loss rate, $\dot{M}_\odot$, because these would cause strong deviations from the observed  
$V-H$ and the degree of linear polarisation. 
The best model was achieved with \mbox{$\dot{M}_\star=50\,\dot{M}_\odot$} 
that corresponds to the estimated wind strength during quiescent phases \citep{Augereau&Beust2006}.

Due to strong dust transport from the planetesimal belt towards the star induced by the stellar wind drag, 
the inner disc region is filled with small scattering grains, 
contrary to what was derived previously \citep{Graham2007, Fitzgerald2007}.
Significant deviations between the modelled and observed $V-H$ and the degree of linear polarisation are visible.
The problem may be mitigated by irregularly-shaped grains that have scattering properties 
different from the Mie spheres used in this study.
We find a preference for particles that are weaker polarisers than Mie spheres and have 
polarisation maxima at scattering angles different from 90$^\circ$. 
This again demonstrates the necessity 
to incorporate more realistic light scattering physics in simulations
of resolved debris discs.

Our models support the presence of ice-containing particles of moderate porosity.
As found by a comparison with other works, porous particles 
seem to be ubiquitous in discs around stars of the $\beta$~Pic moving group.

The radial width of the plantesimal belt cannot be constrained tightly. 
Belts with radial extents of 5 and 17\,au are consistent with the observations.
Furthermore, we have addressed an alternative scenario with a
very broad planetesimal belt, extended over 1~--~45\,au.
The belt was assumed to be populated with particles up to 100\,m in size and to have an MMSN-like surface density 
with initial radial index of $-$1.5.
The inside-out evolution of this belt resulted in a rising surface density with distance from the star
and a 1.3\,mm profile as observed by ALMA.
The scenario explains the formation of an outer ring-like planetesimal distribution 
and would be preferred if the presence of planets that 
cleared up the inner disc region are not confirmed in the future.

We have shown that the unresolved central emission at 1.3\,mm, first reported by \cite{MacGregor2013}, 
cannot stem from an inner dust ring alone. 
An inner ring located at $\lesssim$\,3\,au, which is compatible with being unresolved in the ALMA image, 
emits at most 60\% of the observed central emission.
We found a fractional luminosity ratio between inner and outer component of
\mbox{$\approx0.4$} -- a typical value for a two-temperature disc system \citep{Kennedy&Wyatt2014}.  
As suggested in \cite{MacGregor2013}, future ALMA observations at higher resolution
should be able to test the inner ring hypothesis. 
If confirmed, this might argue against the very extended belt scenario, 
based on the assumption that the inner planetesimals have not survived 
the inside-out collisional erosion of the disc.
However, this assumption remains controversial because some models would explain the presence 
of an inner dust ring even after the collisional depletion of an extended outer ring.
For example, the innermost particles of the extended planetesimal belt could be less icy, 
having a higher mechanical strength, and thus, would resist the collisional grinding more strongly.
This increases the chance to detect left-over particles from the depleted, originally extended, planetesimal disc.  
Furthermore, cometary activity may play a crucial role to create hot dust in the inner disc zone.
Altogether, because the central emission cannot be explained by dust emission alone, it 
fully or at least partly derives from the stellar chromosphere.

Our work provides a good indication of the probable generic architecture of the system, characterised by the spatial 
and size distributions of planetesimals and their collisionally-produced dust.
All the models presented here are not unique as 
a full exploration of the parameter space by collisional modelling is impossible. 
However, we verified consistency of the collisional modelling results with a multidimensional 
power-law fitting approach for the radial and size distributions of the dust.
This reduces the chance that our models show rather ``exotic'' solutions.


\begin{acknowledgement}
We thank the reviewer for a speedy and constructive comments that helped to improve the manuscript.

This paper makes use of the following ALMA data: ADS/JAO.ALMA\#2011.0.00274.S and ADS/JAO.ALMA\#2011.0.00142. 
ALMA is a partnership of ESO (representing its member states), NSF (USA) and NINS (Japan), 
together with NRC (Canada) and NSC and ASIAA (Taiwan), in cooperation with the Republic of Chile. 
The Joint ALMA Observatory is operated by ESO, AUI/NRAO and NAOJ.
We thank J.~Rodmann for his support during the proposal writing process.

T.L., A.V.K., and S.W. acknowledge the support by the 
\emph{Deut\-sche For\-schungs\-ge\-mein\-schaft} (DFG) through projects 
\mbox{Lo~1715/1-1}, \mbox{Kr~2164/10-1}, and \mbox{WO~857/7-1}. 
S.E. thanks for financial support from DFG under contract WO\,857/7-1 and from ANR 
under contract ANR-2010 BLAN-0505-01 (EXOZODI).
J.P.M. is supported by a UNSW Vice-Chancellor' s Fellowship.
M.C.W. acknowledges the support of the European Union through ERC grant number 279973.

\end{acknowledgement}


\end{document}